\documentclass[sigconf]{acmart}
\AtBeginDocument{%
  }

\copyrightyear{2026}
\acmYear{2026}
\setcopyright{cc}
\setcctype{by}
\acmConference[CHI '26]{Proceedings of the 2026 CHI Conference on Human Factors in Computing Systems}{April 13--17, 2026}{Barcelona, Spain}
\acmBooktitle{Proceedings of the 2026 CHI Conference on Human Factors in Computing Systems (CHI '26), April 13--17, 2026, Barcelona, Spain}
\acmPrice{}
\acmDOI{10.1145/3772318.3791636}
\acmISBN{979-8-4007-2278-3/2026/04}

\usepackage{multirow}
\usepackage{graphicx}
\usepackage{subcaption}
\usepackage{booktabs}
\usepackage{makecell}
\usepackage{enumitem}
\usepackage{pifont}
\newcommand{\xmark}{\ding{55}} 
\usepackage{color}
\graphicspath{ {./images/} }
        
\newcommand{\quotes}[1]{\textit{``#1''}}

\newcommand{\remove}[1]{}
\newcommand{\rev}[1]{\textcolor{black}{#1}}
\begin{document}


\title{Applying Value Sensitive Design to Location-Based Services: Designing
for Shared Spaces and Local Conditions}

\author{Hiruni Kegalle}
\orcid{0009-0002-9758-1368}
\affiliation{%
  \institution{RMIT University}
  \city{Melbourne}
  \country{Australia}}
\email{hiruni.kegalle@rmit.edu.au}

\author{Flora D. Salim}
\orcid{0000-0002-1237-1664}
\affiliation{%
  \institution{University of New South Wales}
  \city{New South Wales}
  \country{Australia}}
\email{flora.salim@unsw.edu.au}

\author{Mark Sanderson}
\orcid{0000-0003-0487-9609}
\affiliation{%
  \institution{RMIT University}
  \city{Melbourne}
  \country{Australia}}
\email{mark.sanderson@rmit.edu.au}

\author{Jeffrey Chan}
\orcid{0000-0002-7865-072X}
\affiliation{%
  \institution{RMIT University}
  \city{Melbourne}
  \country{Australia}}
\email{jeffrey.chan@rmit.edu.au}

\author{Danula Hettiachchi}
\orcid{0000-0003-3875-5727}
\affiliation{%
  \institution{RMIT University}
  \city{Melbourne}
  \country{Australia}}
\email{danula.hettiachchi@rmit.edu.au}

\renewcommand{\shortauthors}{Kegalle et al.}

\begin{abstract}
Location-Based Services (LBS) such as ride-sharing, accommodation, food delivery, and location-driven social media platforms entangle digital systems with physical spaces, thereby generating impacts that extend beyond users to others who share the same environments. Existing design approaches struggle to address the dual challenge of value tensions that arise in shared physical spaces and the locality-specific contexts in which LBS operate. To respond, \remove{we extend Value Sensitive Design (VSD) by introducing} \rev{we introduce} Location-Aware Value Sensitive Design (LA-VSD), \remove{an extended framework that adds three heuristics} \rev{a domain-specific adaptation of VSD} tailored to the distinctive characteristics of LBS. LA-VSD guides designers \remove{to} \rev{through three heuristics to help} (1) identify and prioritise stakeholders through local space-sharing scenarios, (2) adapt empirical methods to capture values and tensions in context, and (3) \remove{embed values} \rev{support value-aligned interactions} across both digital and physical layers of the service. Through a case study of e-scooter sharing in Melbourne, we demonstrate how LA-VSD enables more grounded, context-aware, and actionable design of LBS.
\end{abstract}

\ccsdesc[500]{Human-centered computing~HCI theory, concepts and models}
\keywords{Value Sensitive Design, Location-based Services, Interview, Direct and Indirect Stakeholders, E-scooter Sharing Service}


\maketitle

\section{Introduction}
Location-Based Services are digital platforms that use real-time location data of users or mobile objects to mediate access to physical resources and services \cite{kupper2005location}. Unlike many digital platforms that operate solely online, LBS are embedded in physical environments, making their impacts inseparable from the spaces where they are deployed. This spatial entanglement generates value tensions that affect both direct stakeholders, who engage with the platform through its interface, and indirect stakeholders, who experience its consequences without using it. For example, accommodation platforms such as Airbnb have been linked to neighbourhood concerns over noise, waste, and the erosion of community ties; a recent survey in the United States reported that only 30\% of respondents believed Airbnb guests cared about their neighbourhood's well-being \cite{anytime_airbnb_2025}. Similar tensions emerge across other LBS, including ride-sharing \cite{henao2019impact, schaller2021can}, food delivery \cite{brown2020impeding, chen2019impact}, location-driven social media \cite{vsmelhausova2022instagram}, and micro-mobility services \cite{antoniazzi2023safety}, where conflicts arise from issues of space sharing and local community impact.

These value tensions have been acknowledged in prior work \cite{vines2013configuring, iversen2012values}, and different design approaches have been proposed to address them \cite{Friedman2013, otuu2025should}. Value Sensitive Design is one such methodology, offering a well-established framework for integrating human values into technology design \cite{friedman2007human, Friedman2013}. It has been applied across domains ranging from healthcare to public infrastructure, and from assistive technologies to artificial intelligence \cite{friedman2004office, 10.1145/1463160.1463176, 10.1145/1753846.1753870, van2020designing, de2020digital}. Yet \remove{most applications remain focused on digital systems, with only limited efforts extending VSD to contexts involving} \rev{limited work has examined how to apply it in domains where values are shaped through continuous interactions in shared} physical spaces. Applying VSD to Location-Based Services raises distinct challenges: (1) because LBS operate through shared physical environments, values and tensions emerge in embodied interactions among diverse stakeholders; (2) their impacts are shaped by locality-specific norms, practices, and regulations. While prior studies have noted these difficulties, there remains a lack of systematic guidance for identifying and incorporating stakeholder values and experiences that are shaped by both spatial and digital dimensions, as well as by context-specific conditions.

In this paper, we address this gap by \remove{extending Value Sensitive Design through the development of three heuristics, which together form the Location-Aware VSD (LA-VSD) framework. These heuristics respond to two defining characteristics of LBS: their operation through shared physical environments and their locality-dependent impacts shaped by norms, practices, and regulations.}\rev{introducing Location-Aware VSD (LA-VSD), a domain-specific enrichment of VSD that facilitates its application in LBS contexts. Rather than proposing an alternative theory, LA-VSD provides practical heuristics that help designers operationalise VSD that are especially consequential in LBS: understanding shared-space interactions, capturing locality-dependent values, and aligning values into digital and physical components of the service.} We demonstrate the application of LA-VSD in the context of e-scooter sharing services, a widely deployed form of LBS that exemplifies the complex interplay between digital infrastructures, spatial environments, and multiple stakeholders. The contributions of this work are:
\setlist{nolistsep} 
\begin{itemize}
    \item \em{Conceptual}: We introduce three heuristics that \remove{extend VSD to address the design challenges posed by} \rev{support the application of VSD to LBS by making explicit} the spatial entanglement and locality-dependent nature of LBS.
    \item \em{Empirical}: We demonstrate the utility of these heuristics through a case study of e-scooter sharing, showing how they guide the identification and prioritisation of indirect stakeholders, the adaptation of empirical methods to capture values and tensions, and the incorporation of stakeholder values into both digital and physical layers of design.
\end{itemize}

\section{Related Work}
We begin this section by reviewing related work on Location-Based Services and their entanglement with physical space, highlighting how they reshape shared environments and affect both direct and indirect stakeholders. We then turn to design approaches in HCI, outlining how different methods from User-Centred Design to Participatory and Inclusive Design prioritise particular stakeholder groups. Next, we discuss Value Sensitive Design, reviewing its applications across domains and methodological developments. Finally, we identify the limitations of VSD when applied to technologies embedded in shared physical spaces and locality-specific settings, motivating the need for the heuristics we propose.

\subsection{Location-Based Services}

Location-Based Services are digital platforms that generate, filter, or deliver information based on the real-time location of users or mobile objects \cite{kupper2005location}. Operating at the intersection of digital and physical infrastructures, LBS leverage geospatial data to provide location-specific services across domains such as home-sharing, food delivery, and navigation. Unlike traditional digital services, LBS are deeply entangled with physical environments, influencing patterns of urban mobility, reshaping commercial spaces, and redefining access to services. 

A growing body of research has examined the spatial dynamics of LBS. For instance, studies of Airbnb in Barcelona \cite{lagonigro2020understanding} and Nanjing \cite{sun2021characteristics} show that rental distributions are shaped not only by urban structure but also by socioeconomic and regulatory contexts. In urban transport, ride-sharing trip data from New York City revealed how consumer surplus varies substantially across neighbourhoods \cite{lam2021geography}. Food delivery services such as Deliveroo and Foodora also reshape urban space by creating delivery zones, deploying geofencing, and managing couriers through GPS tracking \cite{heiland2021controlling}. In China, food delivery activity has been shown to concentrate in dense urban areas, driven by accessibility and built environment factors \cite{wang2021impacts}. Beyond mobility and service provision, social media platforms with location features are now critical drivers of spatial change. For instance, TikTok virality has transformed remote destinations into overcrowded destinations \cite{wengel2022tiktok} and Instagram reshapes commercial and cultural spaces through concentrated tourist flows \cite{boy2017reassembling}.

Because physical spaces are shared, the impacts of LBS extend well beyond direct stakeholders who use these services. Neighbours of Airbnbs, pedestrians negotiating space with e-scooters, and residents in areas with app-redirected traffic experience consequences without necessarily engaging with the platforms themselves. In this way, LBS propagate effects across multiple layers of society, producing both intended and unintended outcomes. Table~\ref{tab:Location-Based Service Platform_examples} summarises examples of LBS, highlighting their direct and indirect stakeholders alongside the types of impacts they generate.

\begin{table*}[h]
    \centering
    \captionsetup{justification=centering}
    \caption{Stakeholders and impact scenarios of LBS\\}
    \label{tab:Location-Based Service Platform_examples}
    \small
    \begin{tabular}{lllp{7cm}l}
        \toprule
         App & Direct Stakeholder & Indirect Stakeholder & Example Scenario & Literature \\
       \midrule
       Short-term rentals & Host, Guest & Neighbours, Hotels,  & Noise disturbances, increased garbage, housing crisis,  & \cite{doi:10.1080/13683500.2019.1669540}, \cite{10.1145/2872427.2874815}\\
       (e.g., Airbnb, &  & Local businesses, &  reduced availability of long-term rentals,  & \cite{doi:10.1080/13683500.2019.1669540}, \cite{doi:10.1080/01944363.2016.1249011} \\
      Booking.com)  &  & Municipal authorities &  increased local tourism, strain on local infrastructure &  \\
        \midrule
        
        Ride share services & Driver, Passenger  & Local residents,    & Increased traffic congestion, competition with traditional   & \cite{wallsten2015competitive}, \cite{WILLIS202194}\\
       (e.g., Uber, Didi) &  & Taxi services,   &  taxi services, impact on parking availability,  air quality, & \cite{schaller2021can}, \cite{henao2019impact}\\
       &  &  Motor vehicle drivers,   & increased accessibility to local business,   & \cite{jin2019uber} \\
       & & Local businesses,&  reduced demand for public transport &  \\
       & & Municipal authorities&   & \\
       
       \midrule
        Micro-mobility   & Service provider, & Pedestrians, Cyclists,  & Conflict over shared road space, pedestrian safety concerns, &\cite{antoniazzi2023safety},\cite{bai2021relationship} \\
      sharing services & Rider & Local residents,   &   clutter from improperly parked scooters/bikes,  & \cite{tokey2022analysis} \\
        (e.g., Lime, Beam)&  & Local businesses, & increased accessibility to local businesses, &  \\
       & & Municipal authorities & reduced carbon emissions, reduced congestion &\\
       
        \midrule
        Food delivery   & Delivery driver,   & Local restaurants,   & Increased traffic, competition with   &  \cite{zou2016empirical}, \cite{chen2022food}\\
       services (e.g., Uber & Customer,& Grocery stores,   & local food businesses, impact on local food  & \cite{chen2019impact},  \cite{ramirez2023street} \\
     eats, DoorDash)  & Restaurant  & Local residents,  & pricing, noise, parking issues, increased  & \cite{chen2017parking} \\
     & & Dine-in customers & access to local food options &  \\

       \midrule
        Social media  & Content creators,   & Local residents, & Increased traffic,  Increased business for local shops,  & \cite{vsmelhausova2022instagram}, \cite{le2018effects}  \\
       platforms (e.g.,  & Tourists & Local businesses, & promotion of local culture and attractions, &  \cite{zaharani2021impact} \\
      Instagram, Explorest) &  & Environmental agencies, &   environmental degradation, disruption of local life,& \\
      & & Municipal authorities & strain on public infrastructure   & \\

        \midrule
        E-commerce platforms & Sellers,   & Local businesses,    & Increased traffic and congestion, packaging waste, &  \cite{xu2024economic}, \cite{steever2019dynamic}\\
        (e.g., Amazon, E-bay)  & Delivery drivers & Local residents,    &  potential decline in small business sales, &  \cite{pourrahmani2021crowdshipping} \\
       & Customers & Municipal authorities &   competition with local stores&\\

        \midrule
        Navigation Apps  & Drivers,   & Local residents,    & Increased traffic in residential areas, disruption of &  \cite{schade2024traffic}, \cite{mocanu2020bucharest} \\
        (e.g., Google maps)  & Pedestrians, & Traffic management     &   local community life,  safety risks in quiet neighbourhoods, &\cite{phuangsuwan2024impact}  \\
       & Businesses &  authorities & increased local businesses &  \\
        \bottomrule

    \end{tabular}
\end{table*}

\subsection{Stakeholder Engagement in System Design}

\begin{figure}[h]
\vspace{-1em}
\centerline{\includegraphics[width=0.9\columnwidth]{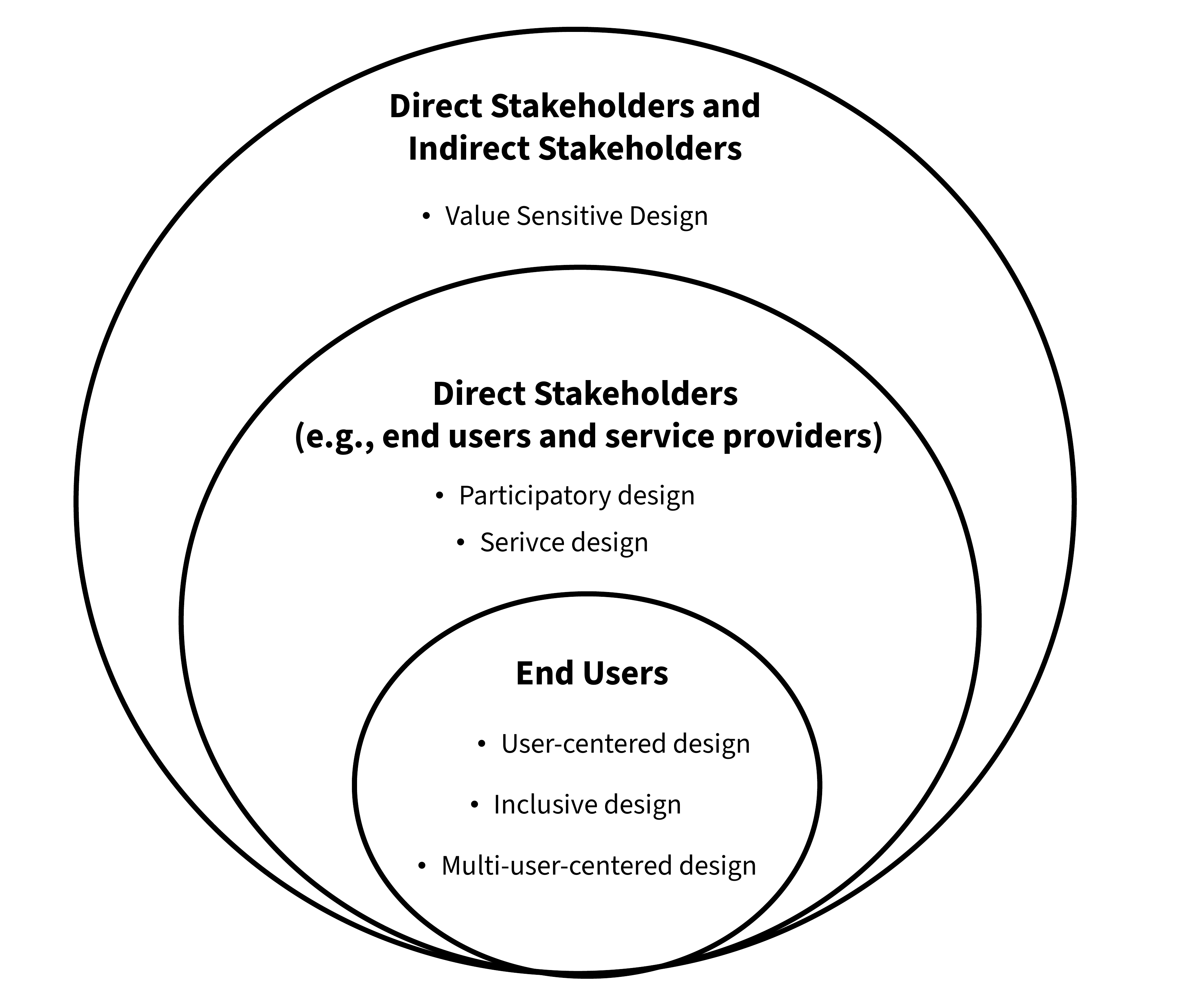}}
\caption{Human-Centred Design approaches categorised by stakeholder focus.}
\label{fig:hcd_lit}
\end{figure}

Human-Centred Design (HCD) \cite{norman1986user,gould1985designing, de2015hci} is a broadly discussed area in Human-Computer Interaction, grounded in the idea that effective design emerges from a deep understanding of people and their needs \cite{auernhammer2020human}. However, within the broader spectrum of HCD, design approaches differ in terms of the stakeholders they prioritise (Figure \ref{fig:hcd_lit}). 

One of the most established frameworks is User-Centred Design (UCD), which primarily focuses on the end user's needs and experiences \cite{10.1145/944519.944525, 10.1145/1047671.1047677}. While effective in improving usability, UCD's narrow focus on direct users can overlook broader social, ethical, and systemic implications. Inclusive Design broadens this scope by aiming to make products and services accessible to a wide range of users \cite{clarkson2013inclusive}. Within this approach, age and disability are often highlighted as key considerations. Recognising the collaborative nature of many technologies, the Multi-User-Centred approach has been proposed to design for systems where multiple users must simultaneously find value in interaction \cite{doi:10.1080/1463922X.2023.2166623}. 

Beyond individual users, Service Design explicitly considers both customers and service providers (e.g., organisations) as core stakeholders \cite{stickdorn2012service}. Participatory Design (PD) goes further by involving users directly as co-designers, embedding democratic principles in the design process \cite{van2014participatory}. Yet in practice, participatory design often remains confined to organisational settings such as workers and management, and rarely extends to wider groups indirectly affected by technology \cite{borning2004designing}. 

To bridge this gap and extend beyond direct stakeholders, \citet{friedman1996value} introduced Value Sensitive Design, an approach that incorporates the values of both direct and indirect stakeholders.

\subsection{Value Sensitive Design}

Value Sensitive Design is a theoretically grounded approach that integrates human values systematically into technology design. In this context, ``\textit{value}'' refers to what is important to people in their lives, with a particular focus on ethics and morality \cite{davis2006value}. VSD emphasises that values emerge through interactions between people and technology, rather than being inherently embedded in technology or solely dictated by social forces \cite{davis2006value}. The framework involves three types of investigation—conceptual, empirical, and technical \cite{friedman2002value} which support designers in identifying stakeholders, their values and embedding these values into technological systems.

\begin{table*}[t]
    \centering
    \captionsetup{justification=centering}
    \caption{Summary of Prior Work on VSD in Physical and Contextual Applications. \\}
    \label{tab:vsd_lit_summary}
    \small
    \begin{tabular}{p{3cm}lp{7cm}cc}
        \toprule
         Focus Area & Study  & Summary of Contributions &  \multicolumn{2}{l}{\textbf{Guidelines} for applying VSD in}  \\
        \cmidrule(lr){4-5}
        & & & physical space & local contexts \\
        \midrule
         \multirow{3}{3cm}{VSD's potential in physical environment and public-facing technology} & \cite{friedman2006watcher} & Applied VSD to augmented window technology, revealing value tensions between office workers and public space users. & \xmark  & \xmark\\
        \cmidrule(lr){2-5}
        & \cite{friedman2008laying}  & Applied VSD in urban development through a simulation web tool enabling stakeholders to select value-aligned indicators. &  \xmark & \xmark\\

        \cmidrule(lr){2-5}
        & \cite{umbrello2022designing} & Applied VSD to explore value integration in autonomous vehicle decision-making algorithms.   &  \xmark & \xmark\\

        \cmidrule(lr){2-5}
        & \cite{jaljolie2023evaluating} & Applied VSD to OpenStreetMap to address ethical concerns such as privacy in digital mapping.  &  \xmark & \xmark\\
        
        \midrule
         \multirow{3}{3cm}{Context-specific nature of values and limitations of VSD to adopt them}& \cite{10.1145/1518701.1518875} &  \multirow{2}{7cm}{Critiqued VSD's fixed-value approach and lack of empirical guidance.} & \xmark & Re-ordered  \\
            & &  &  & VSD phases.\\
        
        \cmidrule(lr){2-5} 
        &\cite{10.1145/2207676.2208560} & \multirow{2}{7cm}{Critiqued VSD's reliance on universal values, researcher authority, and applicability of value list.} &  \xmark & Acknowledge variability,  \\
        & & & & increase agency, transparency \\
         
        \cmidrule(lr){2-5} 
        & \cite{10.1145/1958824.1958836}  & Demonstrated culturally situated values in technology use, highlighting the need for flexibility in VSD.  & \xmark & \xmark \\
        \bottomrule
    \end{tabular}
\end{table*}

\subsubsection{Applications of VSD} \label{VSD_lit}
VSD has been widely adopted in HCI, with applications spanning domains such as healthcare \cite{schikhof2008under,strikwerda2022value}, public transport \cite{watkins2013attitudes}, education \cite{deibel2011understanding}, and safety technologies for vulnerable populations \cite{10.1145/1978942.1979191}. These studies demonstrate VSD's ability to systematically elicit and negotiate stakeholder values, revealing tensions and expanding design considerations beyond usability. VSD has also been applied in Industry 5.0 contexts \cite{longo2020value}, AI for social good \cite{umbrello2021mapping}, and emerging nano-bio-info-cogno technologies \cite{umbrello2020imaginative}, underscoring its adaptability across diverse domains.  

\subsubsection{Methodological Developments and Critiques}
Beyond domain-specific applications, researchers have contributed to the methodological development of VSD, mostly looking at the challenge of resolving value conflicts. Outlining strengths and weaknesses of VSD, \citet{davis2015value} introduced a set of heuristics to enhance its practical implementation. Their work emphasised the need for translating high-level values into concrete design elements that can be embedded into software, using value-based argumentation to resolve stakeholder conflicts, and designing for responsibility by recognising that technological artifacts delegate moral decisions to users. \citet{kozlovski2022parity} examined the challenge of value conflicts, that cannot be measured against each other on a single scale. They critiqued VSD's existing conflict resolution methods and introduced parity as an alternative approach. \citet{manders2011values} argued that VSD lacks a rigorous ethical foundation, pointing to unclear stakeholder identification methods, ambiguity in empirical study integration, and the absence of a structured ethical framework for resolving value conflicts. They proposed Value Conscious Design (VCD) as an enhancement to VSD. Together, these efforts show that VSD has inspired adaptations, critiques, and complementary frameworks.

\subsubsection{Limitations of VSD in Shared Physical Spaces} 
Several studies have explored VSD's potential in shaping physical environments and public-facing technologies. One example is ``augmented window'' study \cite{friedman2006watcher}, which replaced traditional office windows with plasma screens projecting real-time footage of a nearby public plaza. The design highlighted a tension between office workers, who benefitted from enhanced workspace ambience, and people in the plaza, whose images were displayed without explicit consent. While this work illustrates how VSD can surface value conflicts, it does not capture the complexities of multiple stakeholders simultaneously sharing and negotiating the same space.  

Urban development provides another context where VSD has been applied. \citet{friedman2008laying} involved planning officials, advocacy groups, and residents in modelling trade-offs among sustainability, affordable housing, walkability, and business expansion. Their system allowed stakeholders to select indicators aligned with their values, supporting deliberation in decision-making. However, this approach remained at the level of planning support rather than offering concrete design guidance for technologies embedded in everyday shared spaces. 

More recent applications to emerging technologies face similar challenges. In autonomous vehicles, \citet{umbrello2022designing} examined how moral values could be \remove{embedded} \rev{supported} in algorithmic decision-making, focusing on matrix-based approaches to program model behaviour. Although useful for highlighting ethical dimensions of AV design, the work remained centred on computational modelling rather than the lived experience of multiple stakeholders navigating roads together. Similarly, \citet{jaljolie2023evaluating} applied VSD to OpenStreetMap, identifying ethical concerns around privacy and data reliability. Yet the focus remained within the digital domain of mapping, without extending to the contested interactions that arise when maps shape how physical space is shared and navigated.  

Taken together, these studies show that VSD has been used in physical or public-facing technologies, but in highly domain-specific ways. What remains missing are generalisable guidelines for designers to systematically address stakeholder values in shared environments (see Table \ref{tab:vsd_lit_summary}). As LBS expand across multiple domains, the need for such guidance becomes essential.

\subsubsection{Limitations of VSD in Addressing Context-specific Values}
A second limitation of VSD lies in its treatment of values as universal or pre-defined rather than contextually situated. Several works have critiqued this tendency and highlighted how it constrains the discovery of values that emerge from lived, situated practices.  

\citet{10.1145/1518701.1518875} argued that VSD's reliance on fixed value classifications limits its ability to capture context-specific values. They noted that limited methodological guidance for empirical investigations hinders identifying values in real contexts. Based on three case studies, they suggested inverting VSD by placing empirical work before conceptual framing so values arise from practice, not predefined lists.

\citet{10.1145/2207676.2208560} critiqued VSD's focus on universal values, researcher authority in defining ``what matters'', and the applicability of its value lists. They recommended acknowledging cultural variability, contextualising heuristics to socio-cultural settings, amplifying participant agency, and making researcher judgments explicit.  

Finally, research on cross-cultural technology use illustrates the consequences of these limitations. For example, \citet{10.1145/1958824.1958836} examined technology use in long-distance Arabic relationships and showed that values such as privacy, trust, and intimacy were deeply culturally situated. Their findings emphasised the need for VSD to remain flexible enough to account for diverse cultural perspectives, particularly in cross-cultural design contexts where value assumptions cannot be easily generalised.  

Taken together, while these studies underscore the importance of recognising context-specific and culturally grounded values in VSD, they primarily offer critiques and high-level recommendations rather than methodological steps. Systematic guidance for designers to elicit, adapt, and integrate such values into the design process remains missing.  

In summary, VSD provides a strong foundation for systematically integrating values into technology design. Yet, two recurring limitations emerge from prior work. First, there is little methodological guidance for addressing value tensions that arise in shared physical spaces, where multiple stakeholders interact. Second, VSD offers limited mechanisms for capturing context-specific or culturally situated values, often relying instead on universal classifications or researcher-defined priorities. While existing studies provide valuable domain-specific insights and high-level recommendations, they fail to offer generalisable guidance that designers can apply across diverse contexts. Table \ref{tab:vsd_lit_summary} summarises prior VSD applications focused on physical space and context-specific values, underscoring the absence of such generalisable guidelines.

\section{Location-Aware Value Sensitive Design}
    \subsection{Heuristic Generation}
\rev{We follow the methodological approach used by \citet{davis2015value}, who developed VSD heuristics through conceptual synthesis by reviewing VSD applications, critiques, and case studies to identify gaps and articulate guiding principles.
As outlined in Section \ref{VSD_lit}, we systematically examined prior applications (e.g., \cite{watkins2013attitudes}), adaptations (e.g., \cite{umbrello2021mapping}), and critiques (e.g., \cite{kozlovski2022parity}) of VSD alongside literature on LBS. Through this review, we identified two operational challenges in applying VSD to LBS: (a) the absence of generalisable guidance for addressing stakeholder values that emerge through interactions in shared physical environments, and (b) limited methodological support for eliciting, adapting, and integrating context-specific values shaped by local socio-spatial conditions.}

\rev{We synthesised these gaps into an initial set of candidate heuristics within the tripartite VSD methodology. Then we conducted two rounds of critical author discussions, in which the lead author presented the emerging heuristic formulations while the co-authors offered conceptual refinements, and challenged underlying assumptions to ensure rigour and coherence. Finally, we consolidated those into the three heuristics. These heuristics do not claim to expand the theoretical scope of VSD, rather, they make the contextual considerations explicit in LBS design practise, providing structured guidance for operationalising VSD in this domain.}

\begin{figure*}[htb]
  \centering
  \includegraphics[width=0.8\textwidth]{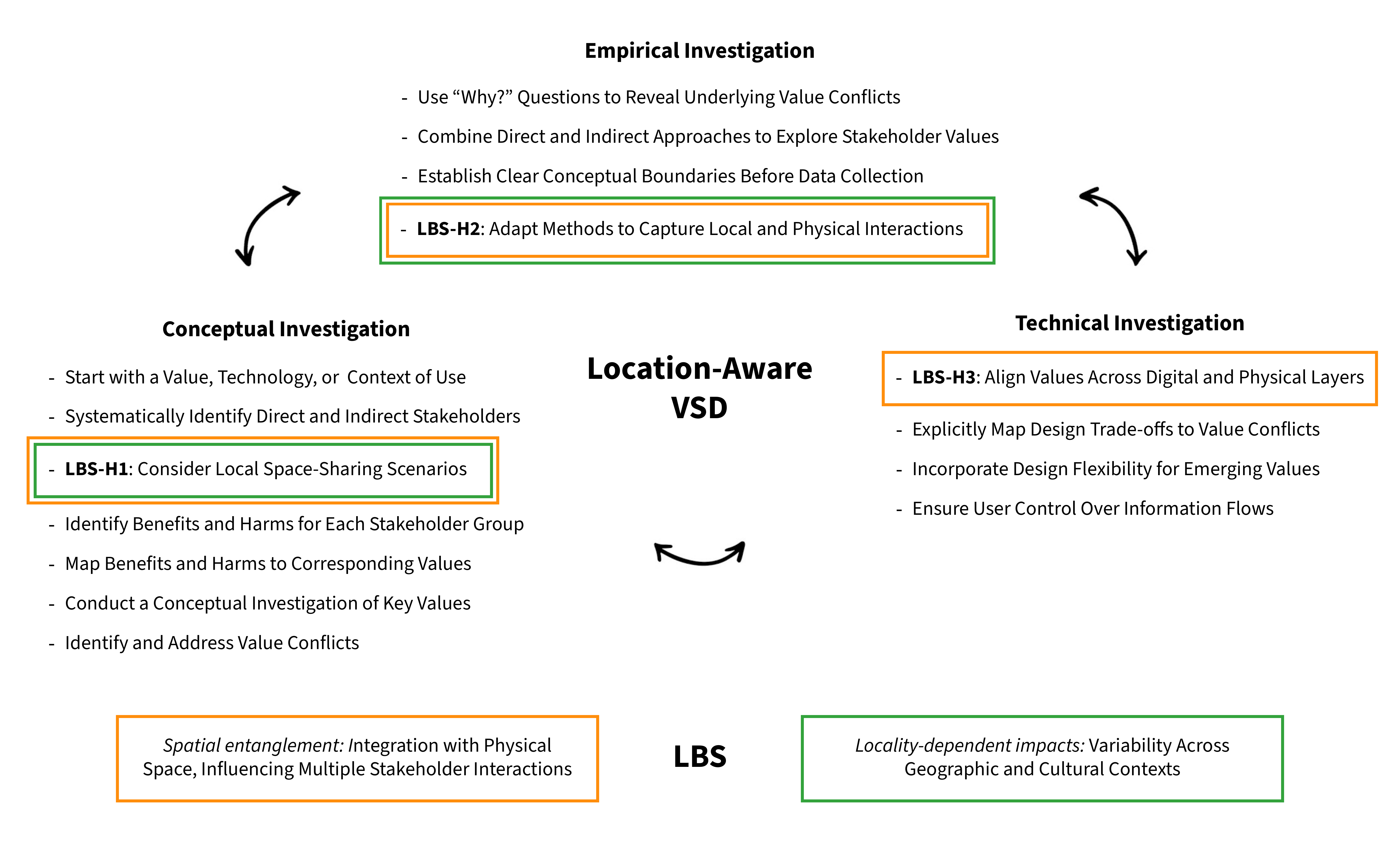}
  \caption{Overview of the Location-Aware Value Sensitive Design framework. The VSD heuristics established by \citet{Friedman2013} are shown alongside the three new heuristics proposed in this paper, which are highlighted in rectangles.}
  \label{fig:extended_framework_new}
\end{figure*}

\remove{The gaps identified in prior work motivated us to extend VSD, addressing the distinctive characteristics of LBS. We propose three heuristics that highlight how values are shaped by the spatial entanglement and the locality-dependent nature.} As seen in Figure \ref{fig:extended_framework_new}, \textit{LBS-H1} and \textit{LBS-H2} reflect both \remove{these} \rev{spatial entanglement and the locality-dependent nature} dimensions, and \textit{LBS-H3} foregrounds spatial entanglement. The heuristics are designed as general guidelines that can be applied across diverse LBS contexts.

\subsection{Heuristics for Conceptual Investigations}

The conceptual investigation phase of VSD seeks to establish a foundation for design by asking key questions: \textit{Who are the stakeholders? What values are implicated? Where do value tensions arise? How might trade-offs among competing values be navigated?} \cite{friedman2002value}. This phase involves a series of steps that guide designers in identifying stakeholders and their values.  

In the following section, we provide a concise overview of the established steps proposed by \citet{friedman2002value}, alongside extended explanations of the \rev{domain-specific} heuristics we introduce. 

        \subsubsection{Start with a Value, Technology, or Context of Use}
        The investigation can begin by focusing on a key value (e.g., privacy, trust, safety), a specific technology, or the context in which the technology is used. This starting point helps frame the investigation and identify relevant stakeholder concerns.
        \subsubsection{Systematically Identify Direct and Indirect Stakeholders}
        Direct stakeholders interact with the technology or its output, while indirect stakeholders are impacted by it without direct interaction. Within these categories, multiple subgroups may exist, and an individual may belong to more than one stakeholder group.
        \subsubsection{LBS-H1: Consider Local Space-Sharing Scenarios} 
        Identifying direct stakeholders, those who directly interact with the technology (service/ platform) is relatively straightforward. However, identifying indirect stakeholders, who may not directly interact, but are affected by its features or outputs is more challenging. Without having a proper understanding of the physical interactions and locality-based nuances, there is a risk of missing out important indirect stakeholder groups. \rev{While VSD supports stakeholder analysis, in LBS, indirect stakeholder relationships are shaped primarily through situated encounters in shared physical environments. These spatial dynamics are central to how stakeholder values emerge. LBS-H1 therefore foregrounds these spatial relations, providing designers with a structured way to operationalise stakeholder identification, without overlooking stakeholder values in the context of LBS.}
        
        To address this challenge, designers should begin by mapping the physical spaces influenced by the platform (e.g., shared roads, footpaths, parks, neighbourhoods) and identifying the groups who occupy or move through them. Space-sharing scenarios can be used to conceptualise how platform's features and outputs might impact them. 
        
        To make these scenarios sensitive to local variation, designers must also draw on locality-specific data. Sources such as municipal reports, community media, research literature, and open datasets can surface cultural norms, regulatory contexts, and community concerns. Incorporating these perspectives ensures that conceptual investigations capture both the spatial entanglement of digital platforms and the locality-dependent nature of stakeholder values.  
        \subsubsection{Identify Benefits and Harms for Each Stakeholder Group}
        Once stakeholders are identified, it is essential to systematically assess how they benefit or harmed by the system. Priority should be given to strongly affected indirect stakeholders and large groups experiencing moderate effects.
        \subsubsection{Map Benefits and Harms to Corresponding Values}
        Identifying benefits and harms enables the recognition of corresponding values. Some mappings are direct, while others are complex and may involve multiple values.
        \subsubsection{Conduct a Conceptual Investigation of Key Values}
        Once values are identified, reviewing relevant literature helps refine their definitions and provides criteria for assessing them empirically. Philosophical and ontological discussions can offer insights into how these values should be incorporated into system design.
        \subsubsection{Identify and Address Value Conflicts}
        Values often come into conflict, and these should be examined not as binary choices but as constraints within the design space. Stakeholder discussions can help in finding balanced solutions.

    \subsection{Heuristics for Empirical Investigations}
    In empirical investigation phase, conceptual ideas are validated and refined through stakeholder engagement \cite{vsd_article}.
        \subsubsection{Use ``Why?'' Questions to Reveal Underlying Value Conflicts}
        When stakeholders express a judgment about a technology or design, asking ``Why?'' can help uncover the underlying reasons for their perspective. A follow-up ``Why?'' question can further clarify the values at play and potential conflicts between different stakeholder concerns.
        \subsubsection{Combine Direct and Indirect Approaches to Explore Stakeholder Values}
        While directly asking stakeholders about a concept (e.g., trust, privacy, or informed consent) can provide insights, it may not always capture the full complexity of their reasoning. Researchers can use alternative approaches such as presenting hypothetical scenarios or discussing common real-life situations to encourage participants to engage in reasoning about the topic in a more natural way.
        \subsubsection{Establish Clear Conceptual Boundaries Before Data Collection}
        Before conducting empirical investigations, researchers need to define key concepts and their scope. Clearly conceptualising the topic ensures that discussions are structured in a way that engages participants in meaningful reasoning. This helps researchers to avoid overly broad or vague discussions. 
        \subsubsection{LBS-H2: Adapt Methods to Capture Local and Physical Interactions}
        Conventional methods such as interviews, surveys, and observations are valuable for studying stakeholder values. But in the context of LBS those methods must be adapted to capture how values arise in shared spaces and are shaped by local contexts. Without such adaptation, designers risk overlooking the dynamics that make LBS distinctive. \rev{While VSD encourages empirical engagement with stakeholders, standard data-collection methods might fail to surface the nuanced, context-dependent value tensions that arise from interactions in public spaces. In LBS, these tensions are situational experiences shaped by the co-presence of multiple stakeholders, and locality-specific features. LBS-H2 therefore operationalises VSD's empirical commitments by making the adaptation of methods to spatial and local conditions an explicit and necessary step.}
        
        For example, interview and survey protocols should explicitly address how the technology/service affects the use of public spaces, and how stakeholders perceive sharing those spaces with others. Observational and in-the-wild studies should be situated in environments where the service is actively in use, so that value tensions become visible in practice. Workshop activities should embed locality by simulating scenarios such as differing regulatory contexts enabling participants to surface trade-offs and context-specific concerns.  

        By adapting established methods in this way, empirical investigations can generate richer insights into how LBS reshape practices and how stakeholder values emerge from both spatial entanglement and locality-dependent variation.

    \subsection{Heuristics for Technical Investigations}
    The technical investigation focuses on designing systems that support the values identified during the conceptual and empirical phases \cite{vsd_article}.

        \subsubsection{LBS-H3: Align Values Across Digital and Physical Layers}
        LBS are hybrid systems, operating simultaneously through digital infrastructures (apps, platforms, algorithms) and physical infrastructures (vehicles, roads, buildings, parking zones). A common challenge for designers is that stakeholder values are realised differently across these layers. While digital design may foreground values such as privacy, transparency, or usability, the same values can manifest quite differently in physical contexts. For instance, usability may refer to ease of navigation in a mobile app, but also to the accessibility of vehicles and infrastructure. \rev{Although VSD supports attention to context, applying it to LBS requires additional focus because digital–physical dual alignment is a primary feature in these systems. LBS-H3 therefore operationalises VSD for the LBS domain by making the alignment of values across digital and physical layers a deliberate and explicit part of the design process.}

        To address this, designers must ensure that values identified through conceptual and empirical investigations are \remove{embedded} \rev{reflected} in both digital and physical components. 
        Digital features such as algorithms and interfaces should be tailored to reflect stakeholder values adapted to local regulations, cultural expectations, and accessibility needs. At the same time, physical components including vehicles, buildings, and the public spaces they occupy must be designed to embody and reinforce same values. In this way, the heuristic adapt VSD beyond interfaces and algorithms, positioning it as a framework that also informs infrastructures and the urban environments where LBS operate.  
        \subsubsection{Explicitly Map Design Trade-offs to Value Conflicts}
        Technical systems often need to balance multiple, sometimes conflicting, stakeholder values. It is helpful to explicitly map design trade-offs to the value conflicts they create and consider how different stakeholder groups are impacted.  
        \subsubsection{Incorporate Design Flexibility for Emerging Values}
        Unforeseen value conflicts often arise once a system is deployed. To accommodate evolving stakeholder concerns, system architectures should be designed with flexibility.
        \subsubsection{Ensure User Control Over Information Flow}
        Managing information flow, particularly in systems that collect and disseminate data through embedded sensors, is a critical and often contested issue. Privacy concerns are heightened in ubiquitous computing environments, making it crucial to provide stakeholders with clear control over data-sharing mechanisms.

\section{LA-VSD Case Study: E-scooter Sharing Services}
To demonstrate the applicability of the LA-VSD framework, we conducted a case study of e-scooter sharing services. The case study comprises a series of actions in which the proposed heuristics were applied in practice \rev{during conceptual and empirical investigations}. In this section, we present these actions alongside the insights they generated (see Table~\ref{tab:results_summary}). 

\begin{table*}[h]
    \centering
    \captionsetup{justification=centering}
    \caption{Summary of heuristics, actions, and outcomes. LBS-H1 and LBS-H2 were empirically applied in this study, while LBS-H3 demonstrates proposed design directions. \\}
    \label{tab:results_summary}
    \small
    \begin{tabular}{lp{5.4cm}p{10cm}}
        \toprule
         Heuristic & Actions / Proposals  & Outcomes \\
        \midrule
LBS-H1 & \multirow{2}{5.4cm}{Reviewed a range of sources, including academic literature and policy documents.} & Recognise riders and service providers as direct stakeholders.\\
& & Identify other road users, governing authorities, businesses and public transport services, etc. as indirect stakeholders.\\

\cmidrule(lr){2-3}
& \multirow{2}{5.4cm}{Determined physical spaces impacted by e-scooter activity.}  & Identify footpaths, cycle lanes, and roads and spaces used for riding.  \\
&   & Identify shared places like parks and junctions that are used for parking. \\

\cmidrule(lr){2-3}
& Explored usage patterns from local data. & Find hotspots and affected POIs.\\
&  & Prioritise pedestrians and cyclists among other road users.\\

\midrule
LBS-H2 & \multirow{2}{5cm}{Designed interview questions to elicit accounts of shared space interactions.}  & Frictions of co-presence in shared space.  \\
& & Breakdowns in safety design.\\

\cmidrule(lr){2-3}
 &  \multirow{2}{5.4cm}{Structured interview questions to probe local regulations and infrastructure practices.} & Impacts of local regulatory variations.\\
  & & Cultural practices shape mobility choices.\\

\midrule
LBS-H3 & \remove{Embed} \rev{Align} values through digital layer.  & Integrate context-aware route recommender system into the App.\\
&& Embed footpath-detection algorithms.\\
&& Guide users to designated parking points through the mobile App.\\
&& Display slow-go and no-go zone activation on the scooter interface.  \\

\cmidrule(lr){2-3}
& \remove{Embed} \rev{Align} values through physical layer. & Add on-path signage for e-scooter allowed paths. \\
&& Implement physical indications for designated parking areas.\\
&& Integrate brighter lights for nighttime use and add built-in turn indicators.\\
&& Enhance scooter stability.\\

        \bottomrule
    \end{tabular}
\end{table*}
    \subsection{Conceptual Investigation}
We applied the heuristic \textbf{LBS-H1: Consider Local Space-Sharing Scenarios}, to identify stakeholders through the lens of space-sharing. We first reviewed academic literature \cite{vsucha2023scooter, doi:10.1080/17450101.2021.1967097, hosseinzadeh2021factors,li2022understanding} to understand broad global trends in e-scooter sharing, and then incorporated local sources -- including news articles \cite{Neuron2025MelbourneReport, ABCNews2024EscooterLegalised}, policy documents \cite{VicParliament2024Escooter}, and accident reports \cite{Pursuit2024EscooterFatalities, ABCNews2023EscooterInjuries} to situate the analysis within the local context. This process informed the identification of riders and service providers as direct stakeholders. Other road users, such as pedestrians, cyclists, and drivers, governing authorities, public transport services, and local businesses, emerged as indirect stakeholders affected by the service.  

\rev{ The next step involved examining} \remove{We then examined} the physical environments influenced by e-scooter service. \remove{Sources indicated that riding activity was concentrated on footpaths, cycle lanes, and roadways, while parking frequently occurred near residential areas, local business areas (e.g., cafés), recreational parks, and public transport stops. } \rev{As suggested in \textbf{LBS-H1}, we conducted a detailed analysis of three months of real-time e-scooter trip data. Each trip record contained trip-start and trip-end coordinates with timestamps. }

\rev{Following approaches established in \citet{kegalle2023footpaths}, we linked this e-scooter trip data to physical environments using two spatial analysis methods. First, we applied machine learning–based spatial modelling to understand which spatial features most strongly influenced trip occurrence. Trip density served as the dependent variable, and the independent variables included a range of spatial attributes, such as land-use types (e.g., residential, commercial, recreational), density of cafés, offices, and public transport stops (tram, bus, and train), population density, and car ownership rates. To understand how these influences varied across temporal contexts, we further trained separate models for weekday vs. weekend and daytime vs. nighttime periods. The resulting feature-importance measures provided insight into which infrastructures and points of interest (POI) categories were most predictive of trip activity.}

\rev{Second, we conducted a buffer analysis to quantify the physical co-location of e-scooter activity with specific environments. Using 10 m buffers around path infrastructures (footpaths, cycle lanes, shared paths), and 60 m buffers around POIs (e.g., office, residence), we calculated the proportion of trip-starts and trip-ends that occurred within immediate proximity to each infrastructure type.}

\rev{Together, these two methods produced a fine-grained understanding of how e-scooter activity is distributed across physical spaces and time, enabling us to identify which stakeholder groups experience the highest spatial co-presence with e-scooters and therefore should be prioritised in our conceptual and empirical investigations.}

\remove{To prioritise stakeholders in our study locality, as \textbf{LBS-H1: Consider Local Space-Sharing Scenarios} suggests, we collected real-time e-scooter trip data over three months. 
Following the approach described by Kegalle et al. \cite{kegalle2023footpaths}, the analysis linked trip data to spatial datasets and revealed usage patterns related to infrastructure and points of interest (POIs). The results showed heavy use of footpaths and cycle lanes, with parking clustering around footpaths. This finding supported us to position pedestrians and cyclists as the most frequently impacted road users. Local accident records and news reports also supported this finding -- stating that recurring safety concerns for pedestrians and cyclists through e-scooter use. In contrast, data analysis results showed limited overlap between e-scooter activity and public transport and local businesses. Therefore, we place those later in our prioritised stakeholder list.  Finally, given that e-scooter sharing was introduced as a city-led trial in the study area, local councils played a central role in regulating and monitoring the service. We identified city councils as a critical indirect stakeholder group.}

\rev{The results of these analyses directly informed how we prioritised stakeholders for this locality. The buffer analysis showed that trip-starts and trip-ends were highly concentrated within 10 m of footpaths and to a lesser extent cycle lanes, indicating that pedestrians and cyclists experience the highest levels of spatial co-presence with e-scooters. In contrast, only a small proportion of trips occurred near public transport stops, business, or office areas. These patterns were reinforced by the machine-learning models: POIs such as tram stops, cafés, and recreational parks showed low or inconsistent feature importance across models and time segments. Temporal modelling further revealed that residential areas were more influential in evenings, and recreational areas spiked only on weekends, indicating that residents near parking hotspots and recreational park users are context-dependent rather than consistently impacted stakeholders. Taken together, these findings allowed us to prioritise pedestrians and cyclists as the most frequently affected stakeholder groups, identify local residents as situationally affected (particularly around high-density parking locations), and place public transport operators and local businesses lower on the prioritisation list due to their limited spatial co-occurrence with e-scooter activity. Councils were also prioritised given their central regulatory role in the trial's implementation and infrastructure management. }

Through this process, LBS-H1 produced a stakeholder map that prioritised pedestrians, cyclists, and councils as the most affected indirect stakeholders in this locality, surfacing value tensions around safety, accessibility, and governance.  

\rev{Without the space-sharing lens, several indirectly affected stakeholder groups—such as local businesses and residents near busy parking areas—would not have been identified. Likewise, without drawing on locality-specific sources, the ways that footpath-riding rules and mixed-use pathways shape stakeholder impacts would have remained obscured. Together, these insights show how LBS-H1 provides a systematic way to surface stakeholders whose experiences emerge from both spatial and local contexts.}

    \subsection{Empirical Investigation}
We employed semi-structured interviews as our empirical method. This approach ensured consistency across stakeholder groups while allowing flexibility for follow-up questions that revealed the reasoning behind participants' perspectives. Guided by \textbf{LBS-H2: Adapt Methods to Capture Local and Physical Interactions}, we developed interview questions that probed how stakeholders experience and negotiate space-sharing with e-scooters in urban environments. For example, participants were asked: \textit{``Can you tell me about the last time you interacted with someone riding a shared e-scooter? Where did this happen, and at what time of day? How did you react, and how did the rider respond?''} To capture routine experiences, we also asked: \textit{``Do you frequently encounter shared e-scooter riders while walking in the CBD?''}  

In addition, the questions were designed to explore contextual factors such as regulations, social norms, current practices, and infrastructure availability that may shape stakeholder perceptions of e-scooter services. For instance, questions included: \textit{``What do you think about the current infrastructure for riding?''}, \textit{``What do you expect from a good cycling experience?''}, and \textit{``Are you aware of the existing e-scooter regulations?''} A full list of interview questions is provided in Appendix \ref{appendix:interview_questions}.  

\rev{This aligns with prior guidance on constructing value-oriented semi-structured interviews proposed by \citet{kahn1999human}, which emphasise designing questions and probes that elicit moral and contextual reasoning.}

The following subsections present details on our participants, study procedure, data analysis process, and findings from the empirical investigation.  

        \subsubsection{Participants}
To ensure participants had relevant local experience, we recruited e-scooter riders, pedestrians, and cyclists from the City Centre (CBD), the area with the highest trip density as per our data analysis results. We assumed that stakeholders in high-use areas would have more frequent interactions with e-scooters, providing richer insights. Recruitment was conducted through community flyers and social media.  

To capture institutional perspectives, we also invited one representative from each local council involved in the e-scooter trial program and one employee from a service provider company. These participants were recruited via email invitations.  

In total, 23 participants took part: 6 e-scooter riders (4 male, 2 female; aged 25–34 = 3, 35–44 = 3), 6 pedestrians (3 male, 3 female; aged 18–24 = 4, 25–34 = 2), 6 cyclists (4 male, 2 female; aged 18–24 = 2, 25–34 = 2, 35–44 = 2), 4 local council representatives, and 1 service provider employee.  

        \subsubsection{Study Procedure}
All participants received an information sheet and consent form prior to their scheduled interview. Interviews were conducted in English via Microsoft Teams and lasted approximately 25–30 minutes. At the start of each session, participants were introduced to the project and consent was reconfirmed before beginning audio/video recording.  

We followed a semi-structured interview guide to maintain consistency while allowing follow-up questions to probe specific topics in depth. Recordings were anonymised using pseudonyms, and transcripts were generated with a third-party transcription service. To enhance reliability, we employed member checking \cite{doi:10.1177/1609406917733847}, sharing transcripts with participants for review and verification of accuracy.  

        \subsubsection{Data Analysis}
We analysed interview transcripts using reflexive thematic analysis \cite{alma9922061070801341, doi:10.1080/2159676X.2019.1628806}. The main researcher first read all transcripts for familiarisation. Two researchers then independently coded five transcripts and developed an initial codebook, which was refined iteratively as further transcripts were coded. To reduce individual bias, the second researcher coded another five randomly selected transcripts \cite{doi:10.1177/1609406917733847}.  

Initial themes were generated by grouping the codes on similarities and patterns. Then, we refined the themes by grouping them into coherent and distinct themes. Finally, themes were defined and given meaningful names. We used Nvivo software to organise the codes in the analysis process.

        \subsubsection{Findings}
Our analysis emerged with four themes, which maps with the two characteristics of LBS. The first two themes reflect spatial entanglement of e-scooter services and the latter two capture locality-dependent impacts. Below, we present the themes.

\textit{\textbf{Frictions of co-presence in shared space.}} Sharing space with others was a recurring source of tension across our interviews. With no dedicated infrastructure, e-scooter riders must navigate footpaths, cycle lanes, and roadways originally designed for other modes. These overlaps created frictions among road users. While, e-scooter riders expressed appreciation for the convenience and efficiency provided by e-scooters, they highlighted several challenges in sharing space. Two riders described the difficulty of manoeuvring through cycling lanes when the same space is served as car parking: \quotes{Once I go to areas where the bike lane is also a parking space, I have to be careful. To get over those parked vehicles and I have to watch over my shoulder just in case there are other cars. Most of the cars give way to bicycles. But, sometimes, there are just people who don't care} [e-scooter rider 3].

Friction was also felt between cyclists and riders. One rider reflected on her shift in perspective when moving from bike to e-scooter: \quotes{bicycles go faster. They were annoyed that we were sharing the paths with them. Because we don't manoeuvre the same way or people are still new to it, so like they're figuring it out. I remember being on a bicycle and being annoyed at e-scooter riders before I was doing it} [e-scooter rider 6]. Cyclists, in turn, highlighted unpredictability from inexperienced riders: \quotes{It might be dangerous, specially those people that are new to e-scooters. [...] When I was cycling, someone was using the e-scooter on the cycle lane.[...] He just did not move on time. So, I had to slow down quickly. I think they can be pretty dangerous. I was pretty shocked and very angry at that person} [cyclist 2].

Pedestrians too voiced frustration when poorly parked scooters blocked their way: \quotes{When I go home late, I found that scooters just laying down, not parked, in the middle of the road where it's hard to cross} [pedestrian 1]. At the same time, riders valued the convenience of free-float parking: \quotes{I try to park [e-scooter] closest to wherever I'm going. So it won't take a long time to just get and leave} [e-scooter rider 5]. These tensions illustrate the double-edged nature of ``dockless'' systems: convenient for users, yet disruptive for others. Service providers acknowledged this tension and described trials of designated parking zones: \quotes{We are investing in finding locations that make sense where vehicles should be parked and labelling them so they're really clear to users. [...] According to the trial, we've had show 98\% compliance and 78\% of riders parking correctly on their first attempt} [service provider]. Further, features like mandatory photo submissions at the end of a ride were also introduced by the service providers to encourage responsible parking and foster community acceptance.  

\textit{\textbf{Breakdowns in safety design.}} Alongside spatial frictions, participants pointed to design limitations—of vehicles, infrastructure, and apps—that left them feeling unsafe. One important concern raised was visibility at night and in poor weather, noting that the built-in lighting was often insufficient. One rider described compensating by adding their own reflective gear and lights: \quotes{At night time, we have to be really cautious with an e-scooter, because we're probably a bit harder to see. I wear reflective gear, and I have extra lights and things like that. Because I acknowledge that the brake light on an E-scooter isn't as visible because it's low to the ground.} [e-scooter rider 6]. This shows how riders adapt to design shortcomings.

Further, the less experienced riders sought safer-feeling spaces, even if this meant riding on footpaths: \quotes{I did not ride it on the bicycle path. Instead, I ride it on the footpaths. Because I feel that's more safe. Even though there might be some people walking in front or walking to me, I will stop and then walk and push the e-scooter instead} [e-scooter rider 1]. This decision reflects the lack of confidence attributed with the e-scooter design—its two-wheel structure and limited stability—which can make cycle lanes feel more precarious than footpaths.

Moreover, cyclists pointed out that they have established social norms/informal practices that help facilitate smoother interactions. But they felt it was still lacking among e-scooter users: \quotes{As cyclists, we do have some hidden rules. [...] when we need to turn, we give a signal to the cyclist behind us by waving our hand. I don't think there are like written rules, but we all do that. I don't know if e-scooters have that kind of things. The problem is, we don't know if they are going straight or they are turning} [cyclist 6]. E-scooter riders themselves explained that this gap is not only about lacking norms but also about the physical design of e-scooters. Unlike bicycles, releasing one hand to signal is unsafe. As per the e-scooter design, it could make them less stable and harder to manoeuvre. Emphasising the difference between two vehicles, an e-scooter rider mentioned: \quotes{bikes are safer. They are naturally more manoeuvrable. The physical design of an e-scooter isn't as smooth as if you needed to brake sharply on a bike. I just ride more carefully. So, if you were to abruptly stop, you could easily flip over} [e-scooter rider 6] 

To ensure safety in shared spaces, the service providers use digital restrictions such as \textit{slow-go} zones (automatic speed reduction) and \textit{no-go} zones (automatic shut-off). However, riders found these features frustrating when their operation was not communicated clearly. One participant described how their scooter abruptly powered down when entering a geofenced area, leaving them stranded without understanding why: \quotes{Once, I got really upset because I rode into a no-riding zone and didn't know it was a no-riding zone. It just shut off on me, and that was really frustrating} [e-scooter rider 6]. This experience highlights a design limitation—not in the safety mechanism itself, but in how it is conveyed to the rider. Without clear notification through the app or the vehicle interface, users are left confused and vulnerable. A more effective design would integrate transparent feedback to ensure that riders understand why their vehicle is slowing or stopping.

Furthermore, participants pointed to the need for better road design to accommodate multiple types of users. Calls for improved infrastructure went beyond scooters to highlight benefits for cyclists and pedestrians as well. As one council officer explained: \quotes{We would need more segregated links, which would also benefit cycling and walking because people would probably use the footpath less for cycling and for scooters. We need more separated trails or shared-use paths with clear directions and enough space to allow this} [council officer 3]. This perspective underscores that safety cannot rely solely on vehicle or service design; it also requires rethinking the built environment to provide clear, dedicated space for different modes of travel.

\textit{\textbf{Impacts of local regulatory variations}} Participants described how local regulations shaped their experience when using e-scooters. One challenge was the maximum speed cap of 20 km/h imposed on shared e-scooters through local regulations. While these limits are introduced as safety measures, riders noted that drivers were often unaware of them, creating moments of tension on the road. As one rider recounted: \quotes{Cars tend to get frustrated if they are behind you because e-scooters are not that quick. They tend to honk their horns and ask you to go faster} [e-scooter rider 2]. These locally varied regulations can unintentionally expose riders to conflict when other road users do not understand or anticipate them.

Moreover, these regulations differed from state to state, adding another layer of complexity. These inconsistencies often left riders unsure about what was permitted, and service providers described the difficulty of ensuring safe and compliant use across jurisdictions. As one service provider explained: \quotes{There are different rules in different states. Some people can run on the footpath, some people can be 16, some people need helmets and all these things. We find when tourists travel here, they don't understand the rules because it's different to those in their own state. That is a challenge we have to face} [service provider]. These fragmented regulations not only confuse users but also create practical challenges for operators tasked with delivering the service.

Helmet requirements provided another example of how regulation shaped adoption. Because helmets are mandatory in our study area, operators embedded shared helmets into the scooter design, unlocked through the app to ensure compliance. While intended to meet regulatory requirements, this design decision introduced new barriers. Helmets were often left outdoors, exposed to weather and dirt, and participants expressed discomfort with sharing them. As one cyclist explained: \quotes{I always want to try e-scooter, but I don't really like sharing the helmet with others because I feel it's a little bit dirty because they just leave the e-scooter outdoors. And then I'm not sure if the helmet will be cleaned up regularly by the organisation of the e-scooter. So I don't really like sharing helmet} [cyclist 6]. Although having helmets embedded in e-scooters is a regulation-driven design choice that improves safety, it has inadvertently discourage potential riders.

\textit{\textbf{Cultural Practices Shape Mobility Choices}} Cultural norms and local transport features strongly shaped decisions about whether to adopt e-scooters. In our study area, the high availability of public transport, walkable streets, and relatively low travel costs (compared with the per-minute pricing of e-scooters) made many pedestrians reluctant to use scooters. As one participant explained: \quotes{I want to use it. But I feel it's too expensive. If I'm going to a place far away, then it's cheaper and easier to take the tram or train. If it is a place nearby, I can use the free tram zone, or I can just walk} [pedestrian 1]. A cyclist echoed similar concerns, contrasting scooters unfavourably with both cycling and public transport: \quotes{Riding a bike, there's no cost. It's free to me. Once, I wanted to try an e-scooter, and then I scanned the QR code on the e-scooter. I checked the price, It is ridiculous! It is even more expensive than the public transport here} [cyclist 6]. 

Cultural expectations around commuting also influenced how scooters were judged. For some cyclists, commuting was not only about mobility but also about building physical activity into everyday life. As one participant explained: \quotes{Cycling is more engaging physically than e-scooter. If I don't go jogging outside or running, then cycling helps me during my day to day life to engage my body in physical activity and burn calories} [cyclist 5]. In this framing, commuting was culturally valued as both transport and exercise, and scooters were seen as falling short.

By applying LBS-H2, we were able to capture the situated experiences of both direct and indirect stakeholders, surfacing values around safety, visibility, and convenience that emerged from how e-scooter services are entangled with physical space and shaped by locality-specific regulations and cultural practices.
   
    \subsection{Technical Investigation}
In the technical investigation phase, stakeholder values identified through the conceptual and empirical stages are translated into system design. In our study, these values were surfaced by applying LBS-H1 and LBS-H2, which highlighted tensions around stakeholder values. Building on these findings, we apply \textbf{LBS-H3: Align Values Across Digital and Physical Layers} to outline \rev{illustrative} design directions that show how \rev{future e-scooter systems could be shaped to better support } stakeholder values \remove{ could be embedded into} \rev{across} both digital and physical components. \remove{ of the e-scooter sharing system.} \remove{ While not implemented in the present study, these proposals illustrate how the heuristic can guide future system design.} \rev{These design directions are conceptual rather than implemented or evaluated; their purpose is to demonstrate how LBS-H3 can orient value-sensitive thinking about system architecture and infrastructure, rather than to serve as validated technical interventions. We see this as a starting point for future work that will co-design, prototype, and assess such value-aligned designs in practice.}

        \subsubsection{\remove{Embed} \rev{Align} Values Through Digital Components}  
In line with \textbf{LBS-H3: Align Values Across Digital and Physical Layers}, we propose several digital design directions that embed stakeholder values into the e-scooter service and enhance its operation. These proposals aim to reduce conflicts, improve safety, and support compliance with local regulations.  

First, to support safer and more predictable interactions between riders and other road users, we propose a context-aware route recommendation system integrated into the service provider's mobile app. This system would generate multiple route options based on contextual factors such as peak and off-peak times, rider experience, and environmental conditions. By surfacing alternatives—similar to existing navigation apps but tuned to scooter-specific factors—riders can select safer, less congested routes, reducing the likelihood of conflict in shared spaces.  

Second, to minimise pedestrian–rider conflicts on footpaths and to support enforcement of no-footpath riding regulations, we suggest embedding footpath-detection algorithms into e-scooters. Riders could be warned when entering footpaths, with gentle nudges to redirect them toward permitted infrastructure. To avoid misclassification in areas where road space is shared between pedestrians and cyclists, the system should be trained on local infrastructure and geospatial data.  

Third, although designated parking points have been introduced in trials, riders are not currently prompted through the mobile app. We propose integrating these points directly into the app interface, providing real-time guidance to the nearest designated location when a rider attempts to end a trip elsewhere. Visualising designated parking in the app would help guide riders toward compliant behaviour while maintaining convenience.  

Finally, to address rider frustration with abrupt slow-go and no-go zone enforcement, we propose displaying these zones directly on the scooter interface. Providing advance feedback when entering regulated areas would reduce surprise and confusion, aligning safety enforcement with transparent communication.  

Together, these proposals demonstrate how digital components can be redesigned to \remove{embed} \rev{support} values of safety, \remove{predictability, and compliance}, \rev{autonomy, and fairness} illustrating the practical application of LBS-H3 in improving the e-scooter system.

        \subsubsection{\remove{Embed} \rev{Align} Values Through Physical Components}  
Inline with \textbf{LBS-H3: Align Values Across Digital and Physical Layers}, we also propose physical design interventions that \remove{embed} \rev{reflect} stakeholder values directly into the built environment and the vehicle itself. 

First, extending and clearly marking dedicated lanes would improve the safety of vulnerable road users. We suggest adding on-path signage specifically for e-scooters, as current shared paths only display pedestrian and cyclist symbols. Providing clear visual cues would reduce uncertainty about where scooters are permitted and help nudge riders toward appropriate spaces. Importantly, visible signage could also promote public acceptance by signalling that scooters are a legitimate part of the transport ecosystem.  

Second, signage should extend to designated parking areas. While trial parking zones have been identified, they are often not physically marked in ways visible to riders. We propose using surface markings or stickers to clearly indicate these areas. As a physical design intervention, this would support compliance by guiding riders to recognised spaces, reinforcing appropriate use of shared environments.  

Third, improvements to the physical design of the vehicle itself could enhance safety. Riders in our study highlighted difficulties with visibility at night and with signalling intentions in traffic. To address these concerns, we propose integrating brighter lights for nighttime use and adding built-in turn indicators, reducing the need for unsafe one-handed gestures. Similarly, enhancing scooter stability—for example, through design modifications to the deck or wheels—would provide riders with greater control and reduce accident risks.  

Together, these physical improvements demonstrate how supporting values into both infrastructure and vehicle design can make e-scooter services safer, more predictable, and more widely accepted, fulfilling the intent of LBS-H3.

\section{Discussion}
We structure our discussion around the importance of the proposed heuristics, interpreting the results in light of these contributions, and connecting them to real-world examples.

\subsection{Importance of LBS-H1}

\subsubsection{Recognise Stakeholders Through Reviewing Sources}
Through reviewing a range of sources, such as academic papers, transport policy documents, and local news reports we observed that physical spaces influenced by e-scooter activity provided a concrete basis for identifying indirect stakeholders. These sources frequently discussed how e-scooters shaped the use of shared paths, roadways, and pedestrian areas, highlighting tensions with pedestrians, cyclists, and drivers who did not directly use the platform but were affected by its operation. For example, policy documents often reported conflicts on footpaths where e-scooters created accessibility challenges for pedestrians \cite{vicroads_escooters}, while local news highlighted concerns from cyclists about sharing narrow lanes \cite{guardian_escooters_2023}. Beyond these stakeholders, sources also pointed to public transport as another stakeholder indirectly influenced by e-scooter activity. In some cities, e-scooters were framed as  ``first and last-mile connectors'' to buses and trains \cite{javadiansr2024coupling, abdollahzadeh2025can, aarhaug2023scooters}. Further, some emphasised benefits for local businesses, with owners attributing improved sales to increased customer access \cite{apollo_escooters_2024,kaabo_escooters_2023}.

These findings indicate that indirect stakeholders in LBS are not simply a peripheral category but can be systematically identified through the lens of shared physical spaces, where indirect stakeholders experience both risks and benefits.  

\subsubsection{Recognise Stakeholders Through Local Data Analysis}
By analysing e-scooter trip data collected in the study area, we were able to prioritise indirect stakeholders based on the spaces most affected by e-scooter activity. The data revealed heavy usage of footpaths and cycle lanes, with parking activity clustering around pedestrian areas. These patterns pointed to pedestrians and cyclists as the most directly impacted indirect stakeholders in this locality. In contrast, the results showed limited overlap between e-scooter activity and sites connected to public transport or local businesses, suggesting that these groups were less directly influenced within this context. The analysis further showed that most activity was concentrated in the CBD, with trip density varying across times of day and days of the week. Impacts on surrounding points of interest (POIs) were also uneven, with higher overlaps during peak commuting hours and weekends compared to quieter periods \cite{kegalle2023footpaths}. 

However, studies in other contexts highlight different impacts. For example, research in the United States cities has shown that electric scooters and dockless bicycles are valued by commuters as effective first–last mile solutions to connect with public transport \cite{grosshuesch2019solving}. More recently, evidence from Louisville, Kentucky demonstrated that e-scooter demand is strongly associated with proximity to local businesses such as shopping malls, restaurants, coffee shops, and bars \cite{karimpour2024estimating}.

\subsubsection{Integrating Both Reviewed Sources and Data}
These findings highlight the value of local empirical trip data for refining stakeholder identification. While review sources presented a broad set of possible indirect stakeholders, the trip data enabled us to determine which groups were most salient in practice for the specific study area. This demonstrates that stakeholder relevance in LBS is highly contingent on local patterns of space use. Pedestrians and cyclists emerged as high-priority stakeholders as their mobility overlapped most intensively with e-scooter operations. Conversely, the absence of strong overlaps with public transport or local business locations underscores that not all theoretically relevant stakeholders carry the same weight in every context.

Taken together, the review of multiple sources and the analysis of trip data illustrate how proposed \textit{LBS-H1:considering local space-sharing scenarios} adds depth and precision to stakeholder identification. Review sources helped us cast a wide net, revealing a broad set of potential indirect stakeholders, including pedestrians, cyclists, drivers, public transport, and local businesses. Data analysis then enabled us to prioritise these groups in the study area, showing that pedestrians and cyclists were most directly affected, while public transport and businesses were less salient in this locality. Moreover, the spatio-temporal patterns in the trip data revealed that stakeholder relevance shifts across times of day, days of the week, and activity zones such as the CBD. Without integrating both perspectives, we would risk either overlooking important stakeholders (if we relied only on data) or treating all stakeholders as equally affected (if we relied only on review sources).

\subsection{Importance of LBS-H2}

\subsubsection{Frictions in Shared Spaces}
Designing interview questions that probed shared-space experiences enabled us to surface frictions of co-presence e-scooter riders. Riders reported navigating bike lanes that doubled as parking spaces, requiring them to watch for opening car doors and inattentive drivers. This finding parallels long-standing cyclist safety concerns \cite{johnson2013cyclists}, highlighting the inadequacy of infrastructure that forces modes into overlapping lanes. Conflicts also arose based on differences in speed and manoeuvrability from cyclists expressing frustration and risks caused by inexperienced e-scooter users. This mirrors findings from multimodal sensing studies, which report that e‑scooter riders struggle to keep pace with faster-moving cyclists, increasing uncertainty in shared lanes \cite{kegalle2025watch}. This shows the challenges of co-locating vehicles with divergent dynamics and skill levels in the same lane. 

Pedestrians described the inconvenience of poorly parked scooters obstructing footpaths, while riders valued the convenience of parking close to destinations, creating tensions with pedestrian accessibility. Prior work has similarly highlighted the accessibility issues caused by e-scooter parking \cite{bennett2021accessibility, james2019pedestrians}, and our findings extend these debates by showing how riders and pedestrians frame parking as a negotiation of values such as \textit{safety} versus \textit{convenience} and \textit{accessibility} versus \textit{efficiency}. Service providers acknowledged these issues and reported trialling designated parking zones, achieving high compliance rates that suggest the potential of infrastructural solutions. This shows how design-mediated governance can partially resolve these tensions by aligning rider convenience with public accessibility.

\subsubsection{Design and Infrastructure Gaps}
The formatted interview questions helped us uncover limitations in vehicle design, gaps in infrastructure allocation, and shortcomings in mobile app features that collectively shaped how participants felt in shared-spaces. Questions about interactions with other road users led e-scooter riders to highlight the limitations of vehicle design. For instance, participants reported that riding in the dark felt unsafe due to lights being too low to be noticed, and inexperienced riders often preferred footpaths because of stability concerns. Cyclists, when asked about interactions, pointed out the absence of established hand signalling practices among e-scooter users; this revealed a mismatch between expectations of shared-space etiquette and what the e-scooter design allows- removing a hand to signal can destabilise the scooter. This concern aligns with prior research showing that indicating a turn by hand is perceived as more difficult and less secure in terms of stability \cite{locken2020impact}. Riders themselves emphasised that scooters are less manoeuvrable and harder to brake sharply than bicycles, underscoring the need for vehicle improvements. 

Highlighting design gaps in the mobile App, riders reported frustration when scooters abruptly powered down without explanation in no-go zones. This reveals a communication gap between the app and the vehicle, where safety mechanisms are undermined by poor feedback, leaving users confused. Our interview questions on infrastructure usage also revealed calls for better road design to accommodate multiple users, with council officers stressing the need for segregated links and clear directions. 

Without formatting interview questions to elicit physical space interactions, these frictions and multi-layered design problem that spans vehicles, mobile App, and the built environment would remain invisible. Traditional interview approaches tend to focus on digital touchpoints such as app usability, booking, or navigation, but such methods overlook the situated conflicts that arise in shared spaces. By deliberately designing questions that probed physical interactions in shared spaces, we were able to reveal frictions that extend beyond—such as the instability of signaling on e-scooters, the inadequacy of infrastructure shared with cars and bicycles, and the accessibility barriers caused by parking practices. This shift demonstrates how LBS-H2 expands empirical methods to capture the lived realities of physical co-presence.

\subsubsection{Influence of Local Regulations}
The way we formulated interview questions helped us uncover how local regulations shaped the use of e-scooters. Participants noted that drivers were often unaware of the maximum speed limit imposed on shared e-scooters, and expressed frustration when scooters could not keep up with traffic. Because speed limits differ between cities, riders highlighted that other road users are often unaware of locally specific regulations. This aligns with prior research showing that, despite broad similarities in e-scooter governance, significant cross-national variations in speed limits and pavement riding rules lead to diverse usage patterns \cite{ventsislavova2024scooters}. Asking participants about their awareness of e-scooter rules revealed how variability across localities created confusion. Making the issue more complex, service providers emphasised the difficulty of ensuring compliance across jurisdictions, as age requirements, footpath riding rules vary significantly between states specially with tourists. 

Further, our interviews revealed how local regulations shape e-scooter design. In our study area, where helmets are mandatory, shared e-scooters were designed with embedded helmets. By contrast, in jurisdictions without helmet mandates, fleets are typically deployed without helmets. This difference is reflected in the U.S. studies that report low helmet use when helmets are not provided at point of hire \cite{sexton2023shared}. However, design alone did not ensure compliance. Our participants expressed reluctance to use shared helmets left outdoors, citing hygiene concerns. Prior Australian research similarly shows that, despite provision, helmet non-use remains significantly higher among shared e-scooter riders than private riders \cite{ssi2024understanding}. These findings highlight how regulations act as design drivers (determining whether helmets are embedded or omitted), but actual adoption is mediated by local practices and user experience.

\subsubsection{Cultural Norms and Practices}
The structure of our interview questions, which probed local infrastructure and transport practices, also revealed how cultural norms and transport features strongly shaped decisions about whether to adopt e-scooters. In our study area, pedestrians and cyclists were reluctant to use scooters given the availability of affordable public transport, walkable streets, and the free tram zone. This highlights how mobility choices are mediated not only by individual preference but by the relative affordability and accessibility of local alternatives. One participant explained that trams and trains were cheaper for longer trips, while walking or the free tram zone was preferable for shorter journeys, showing how e-scooters struggle to compete in contexts with robust, low-cost options. In contrast, a study in Oslo, Norway found that one-third of main-mode e-scooter trips directly substituted public transport, driven by advantages in trip distance, speed, cost, and flexibility compared to transit \cite{fearnley2022factors}. These cases underscore how adoption patterns are deeply connected on local infrastructures and norms. 

Further, our interviews revealed that cultural expectations around commuting also shaped perceptions of e-scooters. Some cyclists valued commuting as a form of daily exercise, framing cycling as physically engaging and health-promoting, while viewing e-scooters as passive. This demonstrates that adoption barriers are not only economic but also tied to cultural values that shape how mobility practices are judged.

Overall, by structuring our interview questions around LBS-H2, we revealed how local regulations (e.g., speed caps, helmet mandates, jurisdictional inconsistencies), cultural norms (such as commuting as exercise), and transport features (including public transport availability, walkability, and relative cost) fundamentally shape the appeal and adoption of e-scooters. By deliberately asking about local regulations and riding practices in context, we uncovered confusion arising from inconsistent rules, hygiene concerns with embedded helmets, and cultural preferences for physical activity. This demonstrates how \textit{LBS-H2: Adapt Methods to Capture Local and Physical Interactions} expands empirical methods to capture the lived realities of physical co-presence, underscoring the need for context-specific strategies to foster safe and acceptable adoption.

\subsection{Importance of LBS-H3}
While the application of LBS-H1 and LBS-H2 surfaced the identification of stakeholders and their values focusing on shared-local spaces, LBS-H3 addresses the next critical step: how to support value-aligned interactions across both digital and physical components of the service. Unlike the earlier heuristics, our contribution here does not rest on empirical findings but on the design guidelines we propose based on stakeholder concerns identified in our studies.

The case of Melbourne banning shared e-scooters \cite{citiestoday2024melbourneban} illustrates why supporting values across only one layer of the system is not effective. Authorities identified potential parking zones exploring usage data and displayed these on the map of mobile App. Yet this digital intervention was largely passive: riders were not given real-time prompts directing them to designated areas when attempting to park elsewhere. Unsurprisingly, footpaths became cluttered, and inappropriate parking behaviours became a major source of public complaint \cite{abc2024escooterban}. Ultimately, these issues contributed to the decision to ban the service. Our design recommendations addresses this by embedding values across both layers: providing riders with real-time digital feedback that directs them to the nearest compliant zone, and reinforcing this through physical signage that makes these areas visible and legitimate. By aligning digital enforcement with physical visibility, systems can move beyond partial measures toward consistent rider practices and broader public acceptance.

Paris shared e-scooters provide another example of why reflecting values across both layers is essential. In response to rising incidents and safety concerns from riders using footpaths, the city introduced regulations prohibiting footpath riding \cite{bbc2019escooterparis}. Yet unsafe practices persisted, and authorities ultimately banned shared e-scooters altogether \cite{ban2023parisescooter}. This case demonstrates that regulation alone is insufficient to change rider behaviour. As we propose in LBS-H3, values such as safety must be embedded in both digital and physical layers: digitally, through footpath-detection algorithms that divert riders from pedestrian areas and context-aware recommendation systems that suggest safer routes; and physically, through on-path signage and other visible cues that make permitted zones clear and socially legitimate in public space.

While our argument for LBS-H3 stresses the importance of supporting values across both digital and physical layers, our analysis also shows that this is not always necessary or even possible. Some values can only be meaningfully supported by a single layer, for example, safety features that depend on vehicle design, such as stability improvements, brighter headlights, or built-in turn indicators, rather than digital augmentation. In such cases, improvements to the scooter itself are more effective than app-based interventions. Conversely, some values can be supported primarily through digital systems, without requiring changes to infrastructure or vehicle form. Acknowledging this nuance strengthens LBS-H3: it is not a rigid prescription of ``both layers always,'' but a heuristic that prompts designers to ask where values are best realised—sometimes through alignment across physical and digital layers, and sometimes through targeted interventions in a single layer.


\remove{Even though we propose an important extension to the} \rev{While we propose a domain-specific adaptation of the} Value Sensitive Design framework that addresses critical challenges in designing location-based services, it has several limitations. \rev{First, the technical investigation presented in Section 4.3 is a conceptual application of LA-VSD rather than implemented or empirically evaluated interventions. The design directions generated through LBS-H3 illustrate how LA-VSD can guide value alignment across digital and physical layers, but within the scope of this work, they were not prototyped, implemented, or tested with stakeholders. Future research should build on this work by co-designing, developing, and empirically evaluating such designs in collaboration with stakeholders.}
\remove{First} \rev{Second}, our framework was verified through a single use case—shared e-scooter services in one locality. While this focus enabled us to foreground how values are shaped by spatial and contextual conditions, it limits transferability. To establish broader applicability, future research should apply the framework to other LBS and examine its performance across diverse geographic, cultural, and regulatory settings. \remove{Second} \rev{Third}, the evaluation of our proposals was bounded by the short-term and context-specific methods we employed; long-term deployments or comparative studies may surface different patterns of stakeholder interaction and value alignment. Finally, as with many interpretive approaches, our findings are influenced by the perspectives of the research team. We sought to mitigate this through reflexive practices and transparency, yet further validation of the framework across additional services, contexts, and localities remains essential.

\section{Conclusion}
In this paper, we propose Location-Aware Value Sensitive Design (LA-VSD), \remove{an extension} \rev{a domain-specific adaptation} of Value Sensitive Design tailored for the unique challenges of location-based services. LA-VSD introduces three heuristics that respond to two defining characteristics of LBS: their spatial entanglement and locality-dependent impacts.

Using the case of e-scooter sharing services, we demonstrated the effectiveness of these heuristics: \textit{LBS-H1: Consider Local Space-Sharing Scenarios}, LBS-H2: \textit{Adapt Methods to Capture Local and Physical Interactions}, and \textit{LBS-H3: Align Values Across Digital and Physical Layers.} Through LBS-H1, we showed how reviewing local sources and analysing usage data supports the recognition of both direct and indirect stakeholders while identifying the physical spaces most impacted by e-scooter activity. LBS-H2 illustrated the value of designing methods that capture shared-space interactions and probe local regulations and practices, revealing tensions in safety, co-presence, and cultural differences in mobility. \remove{Finally, LBS-H3 highlighted how values can be embedded across digital and physical interventions, ranging from algorithmic improvements in routing and parking guidance to infrastructure enhancements such as on-path signage and lighting.} \rev{While LBS-H3 illustrates how values can be aligned across digital and physical layers, our application of this heuristic in the case study was necessarily conceptual. Rather than producing or evaluating concrete design artefacts, we used LBS-H3 to reason through potential design directions that could better support stakeholder values ranging from algorithmic improvements in routing and parking guidance to infrastructure enhancements such as on-path signage and lighting.}

Taken together, these heuristics underscore the importance of designing LBS with sensitivity to local contexts and the diverse stakeholders they affect. Our findings reveal not only the frictions that emerge in shared spaces but also the pathways through which values can be meaningfully addressed. \remove{By extending VSD in this way, we provide researchers and practitioners with a framework for surfacing and embedding values in location-based services that is actionable, context-aware, and oriented toward responsible and inclusive design.}  \rev{Rather than extending VSD theoretically, LA-VSD offers a focused, practical enrichment that supports researchers and practitioners in operationalising VSD within LBS -- providing guidance that is actionable, context-aware, and oriented toward responsible and inclusive design.}

While H3 illustrates how values may be aligned across digital and physical layers, our application of this heuristic in the case study was necessarily conceptual. Rather than producing or evaluating concrete design artefacts, we used H3 to reason through potential design directions that could better support predictability, safety, and communication across the hybrid infrastructures through which LBS operate. This approach aligns with our aim of demonstrating how LA-VSD can structure value-oriented thinking in LBS contexts, but it also means the resulting design suggestions should be understood as illustrative rather than empirically validated. Future work should build on these conceptual insights by co-designing, prototyping, and evaluating such interventions with stakeholders to more fully assess how digital–physical value alignment can be achieved in practice.

\begin{acks}
This research was supported by the ARC Centre of Excellence for Automated Decision-Making and Society (CE200100005), and funded by the Australian Government through the Australian Research Council. 
\end{acks}
\bibliographystyle{ACM-Reference-Format}
\bibliography{sample-base}

@String{Computing = "Computing" }

@String{Computer = "{IEEE} Computer" }

@String{Springer = "Springer-Verlag" }

@book{alma9922061070801341,
author = {Braun, Virginia and Clarke, Victoria},
address = {London},
booktitle = {Thematic Analysis : A Practical Guide},
isbn = {9781473953246},
language = {eng},
publisher = {SAGE Publications Ltd},
title = {Thematic Analysis : A Practical Guide },
year = {2022 - 2022},
}

@article{doi:10.1080/2159676X.2019.1628806,
author = {Virginia Braun and Victoria Clarke},
title = {Reflecting on Reflexive Thematic Analysis},
journal = {Qualitative Research in Sport, Exercise and Health},
volume = {11},
number = {4},
pages = {589-597},
year  = {2019},
publisher = {Routledge},
doi = {10.1080/2159676X.2019.1628806},
URL = {https://doi.org/10.1080/2159676X.2019.1628806}
}

@article{doi:10.1177/1609406917733847,
author = {Lorelli S. Nowell and Jill M. Norris and Deborah E. White and Nancy J. Moules},
title ={Thematic Analysis: Striving to Meet the Trustworthiness Criteria},
journal = {International Journal of Qualitative Methods},
volume = {16},
number = {1},
pages = {1609406917733847},
year = {2017},
doi = {10.1177/1609406917733847},
URL = { https://doi.org/10.1177/1609406917733847}
}

@article{doi:10.1080/17450101.2021.1967097,
author = {Hebe Gibson and Angela Curl and Lee Thompson},
title = {Blurred Boundaries: E-scooter Riders' and Pedestrians' Experiences of Sharing Space},
journal = {Mobilities},
volume = {17},
number = {1},
pages = {69-84},
year  = {2022},
publisher = {Routledge},
doi = {10.1080/17450101.2021.1967097},
URL = {https://doi.org/10.1080/17450101.2021.1967097}
}

@article{hosseinzadeh2021factors,
  title={Factors influencing shared micromobility services: An analysis of e-scooters and bikeshare},
  author={Hosseinzadeh, Aryan and Karimpour, Abolfazl and Kluger, Robert},
  journal={Transportation Research Part D: Transport and Environment},
  volume={100},
  pages={103047},
  year={2021},
  publisher={Elsevier}
}

@article{li2022understanding,
  title={Understanding spatiotemporal trip purposes of urban micro-mobility from the lens of dockless e-scooter sharing},
  author={Li, Hao and Yuan, Zhendong and Novack, Tessio and Huang, Wei and Zipf, Alexander},
  journal={Computers, Environment and Urban Systems},
  volume={96},
  pages={101848},
  year={2022},
  publisher={Elsevier}
}

@online{Neuron2025MelbourneReport,
  author       = {{Neuron Mobility}},
  title        = {Neuron launches new report: Unlocking the potential of rental e-scooters in Melbourne},
  year         = {2025},
  month        = {April},
  day          = {03},
  url          = {https://www.rideneuron.com/neurons-launches-new-report-unlocking-the-potential-of-rental-e-scooters-in-melbourne/},
  note         = {Accessed: 2025-08-28}
}

@online{VicParliament2024Escooter,
  author       = {{Parliament of Victoria}},
  title        = {Helmet mandates, GPS limits and fines coming for e-scooters},
  year         = {2025},
  month        = {July},
  day          = {31},
  url          = {https://www.parliament.vic.gov.au/news/infrastructure/escooter-regulation/},
  note         = {Accessed: 2025-08-28}
}

@online{ABCNews2024EscooterLegalised,
  author       = {{ABC News}},
  title        = {Share hire e-scooters to be permanently legalised across Victoria},
  year         = {2024},
  month        = {July},
  day          = {19},
  url          = {https://www.abc.net.au/news/2024-07-19/share-hire-escooter-permanently-legalised-victoria/104117372},
  note         = {Accessed: 2025-08-28}
}

@online{ABCNews2023EscooterInjuries,
  author       = {{ABC News}},
  title        = {Melbourne doctors call for greater regulation of e-scooters as injuries rise, clogging up hospitals},
  year         = {2023},
  month        = {December},
  day          = {20},
  url          = {https://www.abc.net.au/news/2023-12-20/e-scooter-injuries-rise-as-hospitals-struggle-to-treat-riders/103246314},
  note         = {Accessed: 2025-08-28}
}

@online{Pursuit2024EscooterFatalities,
  author       = {{University of Melbourne}},
  title        = {One in three Australian e-scooter fatalities are children},
  year         = {2025},
  month        = {August},
  day          = {18},
  url          = {https://pursuit.unimelb.edu.au/articles/one-in-three-australian-e-scooter-fatalities-are-children},
  note         = {Accessed: 2025-08-28}
}

@inproceedings{10.1145/2872427.2874815,
author = {Quattrone, Giovanni and Proserpio, Davide and Quercia, Daniele and Capra, Licia and Musolesi, Mirco},
title = {Who Benefits from the "Sharing" Economy of Airbnb?},
year = {2016},
isbn = {9781450341431},
publisher = {International World Wide Web Conferences Steering Committee},
address = {Republic and Canton of Geneva, CHE},
url = {https://doi.org/10.1145/2872427.2874815},
doi = {10.1145/2872427.2874815},
booktitle = {Proceedings of the 25th International Conference on World Wide Web},
pages = {1385–1394},
numpages = {10},
location = {Montreal QC, Canada},
series = {WWW '16}
}

@article{doi:10.1080/13683500.2019.1669540,
author = {Najmeh Hassanli, Jennie Small and Simon Darcy},
title = {The representation of Airbnb in newspapers: a critical discourse analysis},
journal = {Current Issues in Tourism},
volume = {25},
number = {19},
pages = {3186--3198},
year = {2022},
publisher = {Routledge},
doi = {10.1080/13683500.2019.1669540}
}

@article{doi:10.1080/01944363.2016.1249011,
author = {Nicole Gurran and Peter Phibbs},
title = {When Tourists Move In: How Should Urban Planners Respond to Airbnb?},
journal = {Journal of the American Planning Association},
volume = {83},
number = {1},
pages = {80--92},
year = {2017},
publisher = {Routledge},
doi = {10.1080/01944363.2016.1249011},
}

@incollection{friedman2007human,
  title={Human values, ethics, and design},
  author={Friedman, Batya and Kahn Jr, Peter H},
  booktitle={The human-computer interaction handbook},
  pages={1267--1292},
  year={2007},
  publisher={CRC press}
}

@Inbook{Friedman2013,
author="Friedman, Batya
and Kahn, Peter H.
and Borning, Alan
and Huldtgren, Alina",
editor="Doorn, Neelke
and Schuurbiers, Daan
and van de Poel, Ibo
and Gorman, Michael E.",
title="Value Sensitive Design and Information Systems",
bookTitle="Early engagement and new technologies: Opening up the laboratory",
year="2013",
publisher="Springer Netherlands",
address="Dordrecht",
pages="55--95",
isbn="978-94-007-7844-3",
doi="10.1007/978-94-007-7844-3_4",
url="https://doi.org/10.1007/978-94-007-7844-3_4"
}

@inproceedings{friedman2004office,
  title={Office window of the future? Two case studies of an augmented window},
  author={Friedman, Batya and Freier, Nathan G and Kahn Jr, Peter H},
  booktitle={CHI'04 Extended Abstracts on Human Factors in Computing Systems},
  pages={1559--1559},
  year={2004}
}

@inproceedings{10.1145/1463160.1463176,
author = {Friedman, Batya and Hook, Kristina and Gill, Brian and Eidmar, Lina and Prien, Catherine Sallmander and Severson, Rachel},
title = {Personlig integritet: a comparative study of perceptions of privacy in public places in Sweden and the United States},
year = {2008},
isbn = {9781595937049},
url = {https://doi.org/10.1145/1463160.1463176},
doi = {10.1145/1463160.1463176},
booktitle = {Proceedings of the 5th Nordic Conference on Human-Computer Interaction: Building Bridges},
pages = {142–151},
numpages = {10},
location = {Lund, Sweden},
series = {NordiCHI '08}
}

@inproceedings{friedman2008laying,
  title={Laying the foundations for public participation and value advocacy: Interaction design for a large scale urban simulation},
  author={Friedman, Batya and Borning, Alan and Davis, Janet L and Gill, Brian T and Kahn Jr, Peter H and Kriplean, Travis and Lin, Peyina},
  booktitle={Proceedings of the 2008 international conference on Digital government research},
  pages={305--314},
  year={2008}
}

@inproceedings{10.1145/1753846.1753870,
author = {Friedman, Batya and Nathan, Lisa P. and Lake, Milli and Grey, Nell Carden and Nilsen, Trond T. and Utter, Robert F. and Utter, Elizabeth J. and Ring, Mark and Kahn, Zoe},
title = {Multi-lifespan information system design in post-conflict societies: an evolving project in Rwanda},
year = {2010},
isbn = {9781605589305},
url = {https://doi.org/10.1145/1753846.1753870},
doi = {10.1145/1753846.1753870},
booktitle = {CHI '10 Extended Abstracts on Human Factors in Computing Systems},
pages = {2833–2842},
numpages = {10},
location = {Atlanta, Georgia, USA},
series = {CHI EA '10}
}

@article{de2020digital,
  title={Digital platforms and responsible innovation: expanding value sensitive design to overcome ontological uncertainty},
  author={de Reuver, Mark and van Wynsberghe, Aimee and Janssen, Marijn and van de Poel, Ibo},
  journal={Ethics and Information Technology},
  volume={22},
  number={3},
  pages={257--267},
  year={2020},
  publisher={Springer}
}

@incollection{van2020designing,
  title={Designing robots for care: Care centered value-sensitive design},
  author={Van Wynsberghe, Aimee},
  booktitle={Machine ethics and robot ethics},
  pages={185--211},
  year={2020},
  publisher={Routledge}
}

@article{friedman2002value,
  title={Value sensitive design: Theory and methods},
  author={Friedman, Batya and Kahn, Peter and Borning, Alan},
  journal={University of Washington technical report},
  volume={2},
  number={8},
  pages={1--8},
  year={2002}
}

@inproceedings{10.1145/944519.944525,
author = {Jokela, Timo and Iivari, Netta and Matero, Juha and Karukka, Minna},
title = {The standard of user-centered design and the standard definition of usability: analyzing ISO 13407 against ISO 9241-11},
year = {2003},
isbn = {9781450343244},
url = {https://doi.org/10.1145/944519.944525},
doi = {10.1145/944519.944525},
booktitle = {Proceedings of the Latin American Conference on Human-Computer Interaction},
pages = {53–60},
numpages = {8},
location = {Rio de Janeiro, Brazil},
series = {CLIHC '03}
}

@article{10.1145/1047671.1047677,
author = {Mao, Ji-Ye and Vredenburg, Karel and Smith, Paul W. and Carey, Tom},
title = {The state of user-centered design practice},
year = {2005},
issue_date = {March 2005},
publisher = {Association for Computing Machinery},
address = {New York, NY, USA},
volume = {48},
number = {3},
issn = {0001-0782},
url = {https://doi.org/10.1145/1047671.1047677},
doi = {10.1145/1047671.1047677},
journal = {Commun. ACM},
month = {mar},
pages = {105–109},
numpages = {5}
}

@book{stickdorn2012service,
  title={This is service design thinking: Basics, tools, cases},
  author={Stickdorn, Marc and Schneider, Jakob},
  year={2012},
  publisher={John Wiley \& Sons}
}

@article{doi:10.1080/1463922X.2023.2166623,
author = {Sylvain Fleury and Noémie Chaniaud},
title = {Multi-user centered design: acceptance, user experience, user research and user testing},
journal = {Theoretical Issues in Ergonomics Science},
volume = {25},
number = {2},
pages = {209--224},
year = {2024},
publisher = {Taylor \& Francis},
doi = {10.1080/1463922X.2023.2166623},
}

@article{friedman2006watcher,
  title={The watcher and the watched: Social judgments about privacy in a public place},
  author={Friedman, Batya and Kahn Jr, Peter H and Hagman, Jennifer and Severson, Rachel L and Gill, Brian},
  journal={Human-Computer Interaction},
  volume={21},
  number={2},
  pages={235--272},
  year={2006},
  publisher={Taylor \& Francis}
}

@inproceedings{10.1145/1518701.1518875,
author = {Le Dantec, Christopher A. and Poole, Erika Shehan and Wyche, Susan P.},
title = {Values as lived experience: evolving value sensitive design in support of value discovery},
year = {2009},
isbn = {9781605582467},
url = {https://doi.org/10.1145/1518701.1518875},
doi = {10.1145/1518701.1518875},
booktitle = {Proceedings of the SIGCHI Conference on Human Factors in Computing Systems},
pages = {1141–1150},
numpages = {10},
location = {Boston, MA, USA},
series = {CHI '09}
}

@inproceedings{10.1145/2207676.2208560,
author = {Borning, Alan and Muller, Michael},
title = {Next steps for value sensitive design},
year = {2012},
isbn = {9781450310154},
url = {https://doi.org/10.1145/2207676.2208560},
doi = {10.1145/2207676.2208560},
booktitle = {Proceedings of the SIGCHI Conference on Human Factors in Computing Systems},
pages = {1125–1134},
numpages = {10},
location = {Austin, Texas, USA},
series = {CHI '12}
}

@inproceedings{10.1145/1958824.1958836,
author = {Alsheikh, Tamara and Rode, Jennifer A. and Lindley, Si\^{a}n E.},
title = {(Whose) value-sensitive design: a study of long- distance relationships in an Arabic cultural context},
year = {2011},
isbn = {9781450305563},
url = {https://doi.org/10.1145/1958824.1958836},
doi = {10.1145/1958824.1958836},
booktitle = {Proceedings of the ACM 2011 Conference on Computer Supported Cooperative Work},
pages = {75–84},
numpages = {10},
location = {Hangzhou, China},
series = {CSCW '11}
}

@article{wallsten2015competitive,
  title={The competitive effects of the sharing economy: how is Uber changing taxis},
  author={Wallsten, Scott},
  journal={Technology Policy Institute},
  volume={22},
  number={3},
  year={2015}
}

@article{WILLIS202194,
title = {Using "Big Data" to understand the impacts of Uber on taxis in New York City},
journal = {Travel Behaviour and Society},
volume = {22},
pages = {94-107},
year = {2021},
issn = {2214-367X},
doi = {https://doi.org/10.1016/j.tbs.2020.08.003},
url = {https://www.sciencedirect.com/science/article/pii/S2214367X20302027},
author = {George Willis and Emmanouil Tranos}
}

@article{schaller2021can,
  title={Can sharing a ride make for less traffic? Evidence from Uber and Lyft and implications for cities},
  author={Schaller, Bruce},
  journal={Transport policy},
  volume={102},
  pages={1--10},
  year={2021},
  publisher={Elsevier}
}

@article{henao2019impact,
  title={The impact of ride hailing on parking (and vice versa)},
  author={Henao, Alejandro and Marshall, Wesley E},
  journal={Journal of Transport and Land Use},
  volume={12},
  number={1},
  pages={127--147},
  year={2019},
  publisher={JSTOR}
}

@article{jin2019uber,
  title={Uber, public transit, and urban transportation equity: a case study in New York City},
  author={Jin, Scarlett T and Kong, Hui and Sui, Daniel Z},
  journal={The Professional Geographer},
  volume={71},
  number={2},
  pages={315--330},
  year={2019},
  publisher={Taylor \& Francis}
}

@incollection{antoniazzi2023safety,
  title={Safety aspects of e-scooters in urban areas: Preliminary results on citizens' perception, users' behavior and role of pavement},
  author={Antoniazzi, Arianna and Davoli, Elena and Nodari, Claudia and Crispino, Maurizio},
  booktitle={Roads and Airports Pavement Surface Characteristics},
  pages={125--135},
  year={2023},
  publisher={CRC Press}
}

@article{bai2021relationship,
  title={The relationship between E-scooter travels and daily leisure activities in Austin, Texas},
  author={Bai, Shunhua and Jiao, Junfeng and Chen, Yefu and Guo, Jiani},
  journal={Transportation research part D: transport and environment},
  volume={95},
  pages={102844},
  year={2021},
  publisher={Elsevier}
}

@article{tokey2022analysis,
  title={Analysis of spatiotemporal dynamics of e-scooter usage in Minneapolis: Effects of the built and social environment},
  author={Tokey, Ahmad Ilderim and Shioma, Shefa Arabia and Jamal, Shaila},
  journal={Multimodal Transportation},
  volume={1},
  number={4},
  pages={100037},
  year={2022},
  publisher={Elsevier}
}

@article{chen2022food,
  title={Food delivery service and restaurant: Friend or foe?},
  author={Chen, Manlu and Hu, Ming and Wang, Jianfu},
  journal={Management Science},
  volume={68},
  number={9},
  pages={6539--6551},
  year={2022},
  publisher={INFORMS}
}

@inproceedings{chen2019impact,
  title={Impact assessment of food delivery on urban traffic},
  author={Chen, Li-Wen},
  booktitle={2019 IEEE International Conference on Service Operations and Logistics, and Informatics (SOLI)},
  pages={236--241},
  year={2019},
  organization={IEEE}
}

@article{zou2016empirical,
  title={Empirical analysis of delivery vehicle on-street parking pattern in Manhattan area},
  author={Zou, Wei and Wang, Xiaokun and Conway, Alison and Chen, Quanquan},
  journal={Journal of Urban Planning and Development},
  volume={142},
  number={2},
  pages={04015017},
  year={2016},
  publisher={American Society of Civil Engineers}
}

@article{chen2017parking,
  title={Parking for residential delivery in New York City: Regulations and behavior},
  author={Chen, Quanquan and Conway, Alison and Cheng, Jialei},
  journal={Transport Policy},
  volume={54},
  pages={53--60},
  year={2017},
  publisher={Elsevier}
}

@article{xu2024economic,
  title={Economic Analysis of On-Street Parking with Urban Delivery},
  author={Xu, Zhengtian and Sun, Xiaotong},
  journal={Transportation Science},
  year={2024},
  publisher={INFORMS}
}

@article{pourrahmani2021crowdshipping,
  title={Crowdshipping in last mile deliveries: Operational challenges and research opportunities},
  author={Pourrahmani, Elham and Jaller, Miguel},
  journal={Socio-Economic Planning Sciences},
  volume={78},
  pages={101063},
  year={2021},
  publisher={Elsevier}
}

@article{ramirez2023street,
  title={On-street parking for freight, services, and e-commerce traffic in US cities: A simulation model incorporating demand and duration},
  author={Ramirez-Rios, Diana G and Kalahasthi, Lokesh Kumar and Holgu{\'\i}n-Veras, Jos{\'e}},
  journal={Transportation Research Part A: Policy and Practice},
  volume={169},
  pages={103590},
  year={2023},
  publisher={Elsevier}
}

@article{steever2019dynamic,
  title={Dynamic courier routing for a food delivery service},
  author={Steever, Zachary and Karwan, Mark and Murray, Chase},
  journal={Computers \& Operations Research},
  volume={107},
  pages={173--188},
  year={2019},
  publisher={Elsevier}
}

@article{brown2020impeding,
  title={Impeding access: The frequency and characteristics of improper scooter, bike, and car parking},
  author={Brown, Anne and Klein, Nicholas J and Thigpen, Calvin and Williams, Nicholas},
  journal={Transportation research interdisciplinary perspectives},
  volume={4},
  pages={100099},
  year={2020},
  publisher={Elsevier}
}

@article{vsmelhausova2022instagram,
  title={How Instagram users influence nature conservation: A case study on protected areas in Central Europe},
  author={{\v{S}}melhausov{\'a}, Jitka and Riepe, Carsten and Jari{\'c}, Ivan and Essl, Franz},
  journal={Biological Conservation},
  volume={276},
  pages={109787},
  year={2022},
  publisher={Elsevier}
}

@article{le2018effects,
  title={The effects of Instagram on young foreigners vacation choices in Asian countries},
  author={Le, Duong},
  year={2018},
  publisher={Saimaan ammattikorkeakoulu}
}

@article{zaharani2021impact,
  title={The Impact of Micro-Influencer on Brand Image and Purchase Intention in Local Culinary Products on Instagram},
  author={Zaharani, Gita Fitri Rizky and Kusumawati, Nurrani and Aprilianty, Fitri},
  journal={Proceeding B. 6th ICNEM, no},
  pages={11--13},
  year={2021}
}

@article{schade2024traffic,
  title={Traffic jam by GPS: A systematic analysis of the negative social externalities of large-scale navigation technologies},
  author={Schade, Eve and Savino, Gian-Luca and Gunal, Yasemin and Sch{\"o}ning, Johannes},
  journal={PLoS one},
  volume={19},
  number={8},
  pages={e0308260},
  year={2024},
  publisher={Public Library of Science San Francisco, CA USA}
}

@article{mocanu2020bucharest,
  title={Bucharest Drivers' Perception of Navigation Apps and their Impact on Road Traffic.},
  author={MOCANU, Ruxandra},
  journal={GeoPatterns},
  volume={5},
  number={1},
  year={2020}
}

@article{phuangsuwan2024impact,
  title={The Impact of Google Maps Application on the Digital Economy},
  author={Phuangsuwan, Penpim and Siripipatthanakul, Supaprawat and Limna, Pongsakorn and Pariwongkhuntorn, Nuttharin},
  journal={Phuangsuwan, P., Siripipatthanakul, S., Limna, P., \& Pariwongkhuntorn},
  number={2024},
  pages={192--203},
  year={2024}
}

@article{vsucha2023scooter,
  title={E-scooter riders and pedestrians: attitudes and interactions in five countries},
  author={{\v{S}}ucha, Mat{\'u}{\v{s}} and Drimlov{\'a}, Elisabeta and Re{\v{c}}ka, Karel and Haworth, Narelle and Karlsen, Katrine and Fyhri, Aslak and Wallgren, Pontus and Silverans, Peter and Slootmans, Freya},
  journal={Heliyon},
  volume={9},
  number={4},
  year={2023},
  publisher={Elsevier}
}

@inproceedings{kegalle2023footpaths,
  title={Are footpaths encroached by shared e-scooters? Spatio-temporal Analysis of Micro-mobility Services},
  author={Kegalle, Hiruni and Hettiachchi, Danula and Chan, Jeffrey and Salim, Flora and Sanderson, Mark},
  booktitle={2023 24th IEEE International Conference on Mobile Data Management (MDM)},
  pages={255--264},
  year={2023},
  organization={IEEE}
}

@inproceedings{10.1145/1978942.1979191,
author = {Woelfer, Jill Palzkill and Iverson, Amy and Hendry, David G. and Friedman, Batya and Gill, Brian T.},
title = {Improving the safety of homeless young people with mobile phones: values, form and function},
year = {2011},
isbn = {9781450302289},
url = {https://doi.org/10.1145/1978942.1979191},
doi = {10.1145/1978942.1979191},
booktitle = {Proceedings of the SIGCHI Conference on Human Factors in Computing Systems},
pages = {1707–1716},
numpages = {10},
location = {Vancouver, BC, Canada},
series = {CHI '11}
}

@book{deibel2011understanding,
  title={Understanding and supporting the adoption of assistive technologies by adults with reading disabilities},
  author={Deibel, Katherine Nichole},
  year={2011},
  publisher={University of Washington}
}

@book{kupper2005location,
  title={Location-based services: fundamentals and operation},
  author={K{\"u}pper, Axel},
  year={2005},
  publisher={John Wiley \& Sons}
}

@misc{vsd_article, title={What is value-sensitive design?}, url={https://www.techtarget.com/searchcio/definition/value-sensitive-design-VSD}, journal={TechTarget}, publisher={TechTarget}, author={Mary K Pratt}, year={2024}, month={Aug}, note = {Accessed: 2025-01-20}}

@article{auernhammer2020human,
  title={Human-centered AI: The role of Human-centered Design Research in the development of AI},
  author={Auernhammer, Jan},
  year={2020}
}

@article{lagonigro2020understanding,
  title={Understanding Airbnb spatial distribution in a southern European city: The case of Barcelona},
  author={Lagonigro, Raymond and Martori, Joan Carles and Apparicio, Philippe},
  journal={Applied Geography},
  volume={115},
  pages={102136},
  year={2020},
  publisher={Elsevier}
}

@article{sun2021characteristics,
  title={Characteristics and influencing factors of Airbnb spatial distribution in China’s rapid urbanization process: A case study of Nanjing},
  author={Sun, Shijie and Zhang, Shengyue and Wang, Xingjian},
  journal={plos one},
  volume={16},
  number={3},
  pages={e0248647},
  year={2021},
  publisher={Public Library of Science San Francisco, CA USA}
}

@article{lam2021geography,
  title={The geography of ridesharing: A case study on New York City},
  author={Lam, Chungsang Tom and Liu, Meng and Hui, Xiang},
  journal={Information Economics and Policy},
  volume={57},
  pages={100941},
  year={2021},
  publisher={Elsevier}
}

@article{heiland2021controlling,
  title={Controlling space, controlling labour? Contested space in food delivery gig work},
  author={Heiland, Heiner},
  journal={New Technology, Work and Employment},
  volume={36},
  number={1},
  pages={1--16},
  year={2021},
  publisher={Wiley Online Library}
}

@article{wang2021impacts,
  title={Impacts of food accessibility and built environment on on-demand food delivery usage},
  author={Wang, Zhenzhen and He, Sylvia Y},
  journal={Transportation Research Part D: Transport and Environment},
  volume={100},
  pages={103017},
  year={2021},
  publisher={Elsevier}
}

@article{wengel2022tiktok,
  title={The TikTok effect on destination development: Famous overnight, now what?},
  author={Wengel, Yana and Ma, Ling and Ma, Yixiao and Apollo, Michal and Maciuk, Kamil and Ashton, Ann Suwaree},
  journal={Journal of Outdoor Recreation and Tourism},
  volume={37},
  pages={100458},
  year={2022},
  publisher={Elsevier}
}

@article{boy2017reassembling,
  title={Reassembling the city through Instagram},
  author={Boy, John D and Uitermark, Justus},
  journal={Transactions of the Institute of British Geographers},
  volume={42},
  number={4},
  pages={612--624},
  year={2017},
  publisher={Wiley Online Library}
}

@article{clarkson2013inclusive,
  title={Inclusive design: Design for the whole population},
  author={Clarkson, P John and Coleman, Roger and Keates, Simeon and Lebbon, Cherie},
  year={2013},
  publisher={Springer Science \& Business Media}
}

@inproceedings{borning2004designing,
  title={Designing for human values in a urban simulation system: Value sensitive design and participatory design},
  author={Borning, Alan and Friedman, Batya and Kahn Jr, Peter H},
  booktitle={PDC},
  pages={68--71},
  year={2004}
}

@article{van2014participatory,
  title={Participatory design and design for values},
  author={Van der Velden, Maja and M{\"o}rtberg, Christina and Van den Hoven, Jeroen and Vermaas, Pieter E and Van de Poel, Ibo},
  journal={Development},
  volume={11},
  number={3},
  pages={215--236},
  year={2014}
}

@article{friedman1996value,
  title={Value-sensitive design},
  author={Friedman, Batya},
  journal={interactions},
  volume={3},
  number={6},
  pages={16--23},
  year={1996},
  publisher={ACM New York, NY, USA}
}

@phdthesis{davis2006value,
  title={Value Sensitive Design of interactions with UrbanSim indicators},
  author={Davis, Janet Louise Newman},
  year={2006},
  school={Citeseer}
}

@article{watkins2013attitudes,
  title={Attitudes of bus operators towards real-time transit information tools},
  author={Watkins, Kari Edison and Borning, Alan and Rutherford, G Scott and Ferris, Brian and Gill, Brian},
  journal={Transportation},
  volume={40},
  pages={961--980},
  year={2013},
  publisher={Springer}
}

@article{longo2020value,
  title={Value-oriented and ethical technology engineering in industry 5.0: A human-centric perspective for the design of the factory of the future},
  author={Longo, Francesco and Padovano, Antonio and Umbrello, Steven},
  journal={Applied sciences},
  volume={10},
  number={12},
  pages={4182},
  year={2020},
  publisher={MDPI}
}

@inproceedings{schikhof2008under,
  title={Under watch and ward at night: design and evaluation of a remote monitoring system for dementia care},
  author={Schikhof, Yvonne and Mulder, Ingrid},
  booktitle={Symposium of the Austrian HCI and Usability Engineering Group},
  pages={475--486},
  year={2008},
  organization={Springer}
}

@article{strikwerda2022value,
  title={The value sensitive design of a preventive health check app},
  author={Strikwerda, Litska and Van Steenbergen, Marlies and Van Gorp, Anke and Timmers, Cathelijn and Van Grondelle, Jeroen},
  journal={Ethics and Information Technology},
  volume={24},
  number={3},
  pages={38},
  year={2022},
  publisher={Springer}
}

@article{umbrello2020imaginative,
  title={Imaginative value sensitive design: Using moral imagination theory to inform responsible technology design},
  author={Umbrello, Steven},
  journal={Science and Engineering Ethics},
  volume={26},
  number={2},
  pages={575--595},
  year={2020},
  publisher={Springer}
}

@article{umbrello2021mapping,
  title={Mapping value sensitive design onto AI for social good principles},
  author={Umbrello, Steven and Van de Poel, Ibo},
  journal={AI and Ethics},
  volume={1},
  number={3},
  pages={283--296},
  year={2021},
  publisher={Springer}
}

@article{davis2015value,
  title={Value sensitive design: Applications, adaptations, and critiques},
  author={Davis, Janet and Nathan, Lisa P and others},
  journal={Handbook of ethics, values, and technological design: Sources, theory, values and application domains},
  pages={11--40},
  year={2015},
  publisher={Springer Dordrecht}
}

@article{kozlovski2022parity,
  title={Parity and the resolution of value conflicts in design},
  author={Kozlovski, Atay},
  journal={Science and Engineering Ethics},
  volume={28},
  number={2},
  pages={22},
  year={2022},
  publisher={Springer}
}

@article{manders2011values,
  title={What values in design? The challenge of incorporating moral values into design},
  author={Manders-Huits, No{\"e}mi},
  journal={Science and engineering ethics},
  volume={17},
  number={2},
  pages={271--287},
  year={2011},
  publisher={Springer}
}

@article{umbrello2022designing,
  title={Designing AI for explainability and verifiability: a value sensitive design approach to avoid artificial stupidity in autonomous vehicles},
  author={Umbrello, Steven and Yampolskiy, Roman V},
  journal={International Journal of Social Robotics},
  volume={14},
  number={2},
  pages={313--322},
  year={2022},
  publisher={Springer}
}

@article{jaljolie2023evaluating,
  title={Evaluating current ethical values of OpenStreetMap using value sensitive design},
  author={Jaljolie, Ruba and Dror, Talia and Siriba, David N and Dalyot, Sagi},
  journal={Geo-Spatial Information Science},
  volume={26},
  number={3},
  pages={362--378},
  year={2023},
  publisher={Taylor \& Francis}
}

@article{ssi2024understanding,
  title={Understanding nonuse of mandatory e-scooter helmets},
  author={Ssi Yan Kai, Nathalie and Haworth, Narelle and Schramm, Amy},
  journal={Traffic injury prevention},
  volume={25},
  number={5},
  pages={757--764},
  year={2024},
  publisher={Taylor \& Francis}
}

@misc{vicroads_escooters,
  author       = {VicRoads},
  title        = {E-scooters in Victoria},
  year         = {2025},
  note         = {Accessed: 2025-09-02},
  url          = {https://www.vicroads.vic.gov.au/safety-and-road-rules/e-scooters-in-victoria}
}

@article{guardian_escooters_2023,
  author       = {James Norman},
  title        = {‘Bottom of the food chain’: Australia’s e-scooter users just want a safe space to ride},
  journal      = {The Guardian},
  year         = {2023},
  month        = {September},
  day          = {17},
  url          = {https://www.theguardian.com/australia-news/2023/sep/17/bottom-of-the-food-chain-australias-e-scooter-users-just-want-a-safe-space-to-ride}
}

@article{javadiansr2024coupling,
  title={Coupling shared E-scooters and public transit: a spatial and temporal analysis},
  author={Javadiansr, Mohammadjavad and Davatgari, Amir and Rahimi, Ehsan and Mohammadi, Motahare and Mohammadian, Abolfazl and Auld, Joshua},
  journal={Transportation Letters},
  volume={16},
  number={6},
  pages={581--598},
  year={2024},
  publisher={Taylor \& Francis}
}

@article{abdollahzadeh2025can,
  title={Can E-scooters connect first and last-mile of public rail transit? Lessons learned from intercept user survey in Utah},
  author={Abdollahzadeh Kalantari, Hannaneh and Yang, Wookjae and Ewing, Reid},
  journal={International Journal of Sustainable Transportation},
  pages={1--24},
  year={2025},
  publisher={Taylor \& Francis}
}

@misc{apollo_escooters_2024,
  author       = {{Apollo Scooters}},
  title        = {E-Scooters are Revolutionizing City Planning},
  year         = {2024},
  month        = {December},
  day          = {21},
  note         = {Apollo Scooters Blog},
  url          = {https://apolloscooters.au/blogs/news/e-scooters-are-revolutionizing-city-planning}
}

@misc{kaabo_escooters_2023,
  author       = {{KAABO USA}},
  title        = {The Economic Impact of E-Scooters: How Are They Affecting Local Businesses?},
  year         = {2023},
  month        = {July},
  day          = {22},
  note         = {KAABO USA Blog},
  url          = {https://www.kaabousa.com/blogs/blog/the-economic-impact-of-e-scooters-how-are-they-affecting-local-businesses}
}

@article{grosshuesch2019solving,
  title={Solving the first mile/last mile problem: Electric scooters and dockless bicycles are positioned to provide relief to commuters struggling with a daily commute},
  author={Grosshuesch, Kelly},
  journal={Wm. \& Mary Envtl. L. \& Pol'y Rev.},
  volume={44},
  pages={847},
  year={2019},
  publisher={HeinOnline}
}

@article{karimpour2024estimating,
  title={Estimating an e-scooter origin-destination model leveraging Yelp POI data for enhanced urban mobility insights},
  author={Karimpour, Abolfazl and Karimi, Sajjad and Kluger, Robert},
  journal={Environment and Planning B: Urban Analytics and City Science},
  year={2024},
  publisher={SAGE Publications Sage UK: London, England}
}

@article{johnson2013cyclists,
  title={Cyclists and open vehicle doors: Crash characteristics and risk factors},
  author={Johnson, Marilyn and Newstead, Stuart and Oxley, Jennie and Charlton, Judith},
  journal={Safety science},
  volume={59},
  pages={135--140},
  year={2013},
  publisher={Elsevier}
}

@article{kegalle2025watch,
  title={Watch Out! E-scooter Coming Through!: Multimodal Sensing of Mixed Traffic Use and Conflicts Through Riders' Ego-centric Views},
  author={Kegalle, Hiruni Nuwanthika and Hettiachchi, Danula and Chan, Jeffrey and Sanderson, Mark and Salim, Flora D},
  journal={Proceedings of the ACM on Interactive, Mobile, Wearable and Ubiquitous Technologies},
  volume={9},
  number={1},
  pages={1--23},
  year={2025},
  publisher={ACM New York, NY, USA}
}

@inproceedings{bennett2021accessibility,
  title={Accessibility and the crowded sidewalk: Micromobility's impact on public space},
  author={Bennett, Cynthia and Ackerman, Emily and Fan, Bonnie and Bigham, Jeffrey and Carrington, Patrick and Fox, Sarah},
  booktitle={Proceedings of the 2021 ACM Designing Interactive Systems Conference},
  pages={365--380},
  year={2021}
}

@article{james2019pedestrians,
  title={Pedestrians and e-scooters: An initial look at e-scooter parking and perceptions by riders and non-riders},
  author={James, Owain and Swiderski, JI and Hicks, John and Teoman, Denis and Buehler, Ralph},
  journal={Sustainability},
  volume={11},
  number={20},
  pages={5591},
  year={2019},
  publisher={MDPI}
}

@inproceedings{locken2020impact,
  title={Impact of hand signals on safety: Two controlled studies with novice e-scooter riders},
  author={L{\"o}cken, Andreas and Brunner, Pascal and Kates, Ronald},
  booktitle={12th international conference on automotive user interfaces and interactive vehicular applications},
  pages={132--140},
  year={2020}
}

@article{ventsislavova2024scooters,
  title={E-scooters: Still the new kid on the transport block. Assessing e-scooter legislation knowledge and illegal riding behaviour},
  author={Ventsislavova, Petya and Baguley, Thom and Antonio, Josceline and Byrne, Daniel},
  journal={Accident Analysis \& Prevention},
  volume={195},
  pages={107390},
  year={2024},
  publisher={Elsevier}
}

@article{sexton2023shared,
  title={Shared e-scooter rider safety behaviour and injury outcomes: a review of studies in the United States},
  author={Sexton, Emma GP and Harmon, Katherine J and Sanders, Rebecca L and Shah, Nitesh R and Bryson, Meg and Brown, Charles T and Cherry, Christopher R},
  journal={Transport reviews},
  volume={43},
  number={6},
  pages={1263--1285},
  year={2023},
  publisher={Taylor \& Francis}
}

@article{aarhaug2023scooters,
  title={E-scooters and public transport--Complement or competition?},
  author={Aarhaug, J{\o}rgen and Fearnley, Nils and Johnsson, Espen},
  journal={Research in transportation economics},
  volume={98},
  pages={101279},
  year={2023},
  publisher={Elsevier}
}

@article{fearnley2022factors,
  title={Factors affecting e-scooter mode substitution},
  author={Fearnley, Nils},
  journal={Findings},
  year={2022},
  publisher={Findings Press}
}

@online{abc2024escooterban,
  title   = {Melbourne bans hire e-scooters amid safety concerns},
  author  = {{ABC News}},
  year    = {2024},
  month   = {aug},
  day     = {14},
  url     = {https://www.abc.net.au/news/2024-08-14/hire-e-scooter-ban-melbourne-safety-concerns/104219234},
  note    = {Accessed: 2025-09-02}
}

@online{bbc2019escooterparis,
  title   = {Paris bans e-scooters from pavements},
  author  = {{BBC News}},
  year    = {2019},
  month   = oct,
  day     = {31},
  url     = {https://www.bbc.com/news/world-europe-50189279},
  note    = {Accessed: 2025-09-02}
}

@online{citiestoday2024melbourneban,
  title   = {Melbourne announces ban on shared e-scooters},
  author  = {{Cities Today}},
  year    = {2024},
  month   = aug,
  day     = {14},
  url     = {https://cities-today.com/melbourne-announces-ban-on-shared-e-scooters/},
  note    = {Accessed: 2025-09-02}
}

@online{ban2023parisescooter,
  title   = {What Went Wrong with the Paris Referendum On E-Scooters?},
  author  = {{Karl Dickinson}},
  year    = {2024},
  month   = {Mar},
  day     = {5},
  url     = {https://citychangers.org/e-scooters-referendum-paris/},
  note    = {Accessed: 2025-09-02}
}

@misc{anytime_airbnb_2025,
  author       = {Nick Pisano},
  title        = {Do Airbnbs Make Good Neighbors? 2025 Data Reveals Widespread Concerns About Airbnbs in Neighborhoods},
  year         = {2025},
  month   = {July},
  day     = {7},
  note         = {Anytime Estimate Research},
  url          = {https://anytimeestimate.com/research/airbnb-neighborhoods/}
}

@book{norman1986user,
  title={User centered system design; new perspectives on human-computer interaction},
  author={Norman, Donald A and Draper, Stephen W},
  year={1986},
  publisher={L. Erlbaum Associates Inc.}
}

@article{gould1985designing,
  title={Designing for usability: key principles and what designers think},
  author={Gould, John D and Lewis, Clayton},
  journal={Communications of the ACM},
  volume={28},
  number={3},
  pages={300--311},
  year={1985},
  publisher={ACM New York, NY, USA}
}

@inproceedings{de2015hci,
  title={HCI Design Methods: where next? from user-centred to creative design and beyond},
  author={de Haan, Geert},
  booktitle={Proceedings of the European Conference on Cognitive Ergonomics 2015},
  pages={1--8},
  year={2015}
}

@inproceedings{otuu2025should,
  title={How Should We Design Technology With Diverse Stakeholders Who Wish Not to Attend Design Activities Together?},
  author={Otuu, Obinna Ogbonnia and Sahoo, Deepak},
  booktitle={Proceedings of the 2025 CHI Conference on Human Factors in Computing Systems},
  pages={1--14},
  year={2025}
}

@inproceedings{vines2013configuring,
  title={Configuring participation: on how we involve people in design},
  author={Vines, John and Clarke, Rachel and Wright, Peter and McCarthy, John and Olivier, Patrick},
  booktitle={Proceedings of the SIGCHI conference on human factors in computing systems},
  pages={429--438},
  year={2013}
}

@article{iversen2012values,
  title={Values-led participatory design},
  author={Iversen, Ole Sejer and Halskov, Kim and Leong, Tuck W},
  journal={CoDesign},
  volume={8},
  number={2-3},
  pages={87--103},
  year={2012},
  publisher={Taylor \& Francis}
}

@book{kahn1999human,
  title={The human relationship with nature: Development and culture},
  author={Kahn, Peter H},
  year={1999},
  publisher={mit Press}
}

\appendix

\section{Interview Questions}
\label{appendix:interview_questions}

    \subsection{Interview Questions for Riders}
        \subsubsection{Usage and frequency}
        \begin{enumerate}
\item{In a normal week, how often do you ride a shared e-scooter? Can you tell me what areas you travel, on what days of the week and at what times of the day you use a shared e-scooter?}
\item{What form of transport have shared e-scooters replaced in your case?}
\item{Do you see any benefit in transferring from that form of transport? What are they?}
        \end{enumerate}
    \subsubsection{Interactions}  \hfill\\
Thinking about your last shared e-scooter ride, 
    \begin{enumerate}
\item{Can you tell me what is the area/route you travelled? What time of the day it was? }
\item{What was the main reason for choosing a shared e-scooter for that ride?}
\item{What would be the form of transport you use, if shared e-scooters were not available?}
\item{Did you experience any interaction with pedestrians while riding? 
Were you on the footpath? What was their reaction? What was your reaction?}
\item{Did you experience any interaction with cyclists while riding?  
Were you on the cycle lane?  what was their reaction? What was your reaction?}
\item{Did you experience any interaction with other e-scooter riders?  
What was the infrastructure (path) you were using?  what was their reaction? What was your reaction?}
\item{Did you experience any interaction with other motor vehicles?   
Were you on the motor vehicle lane?  what was their reaction? What was your reaction?}
\item{Did you feel unsafe while riding?}
\item{Did you reach a shop/ cafe while you were riding or at the end of the journey?  
           If yes, would it be the same if you used another form of transport for that trip?
           If no, do you see any difference in accessing shops/cafes using e-scooters and other forms of transport?}
\item{Did you use the shared e-scooter to get connected to public transport? 
            If yes, is it the case every time you use a shared e-scooters? 
            If no, did you complete the whole journey using the e-scooter or combined with another form of transport?}
\item{When you are parking your shared e-scooter, did you think about the distractions that might cause to others?}
\item{Is there any other experience with e-scooter sharing you would like to explain?}
\item{Are you aware of the existing e-scooter regulations?}
        \end{enumerate}

    \subsubsection{Values}
        \begin{enumerate}
\item{Tell me about the good things about using a shared e-scooter.  What are the best things about the ride experience?}
\item{Tell me about the bad things/challenges of using a shared e-scooter?   What are the worst things about the ride experience?}
\item{What path makes you feel safe when riding a shared e-scooter? (Cycle lane, main road, shared path, sidewalk)}
\item{What path makes you the most dangerous when riding a shared e-scooter? }
\item{Do you think it would be helpful to have a system that recommends designated parking spots for e-scooters? As a rider, what would you expect from such a system?} 
\item{Do you think it would be helpful to have a system that recommends routes for e-scooters? As a rider, what would you expect from such a system?}
        \end{enumerate}

    \subsection{Interview Questions for Pedestrians}
        \subsubsection{Usage and frequency}
        \begin{enumerate}
\item{In a normal week, how often do you walk in CBD? 
Can you tell me what areas/routes you walk the most and in what days of the week and at what times of the day you used to walk? }
        \end{enumerate}

        \subsubsection{Interactions}
        \begin{enumerate}
\item{Can you tell me about the last time you interacted with someone riding a shared e-scooter? Where did this happen, and what time of day was it? How did you react to the situation, and how did the e-scooter rider respond?}
\item{Do you frequently encounter shared e-scooter riders while you walk in CBD? Are your interactions with them usually the same, or have you had different kinds of experiences?}
\item{Do you feel unsafe sharing the same space with e-scooter riders?}
\item{Can you tell me about the last time you interacted with a ``parked'' shared e-scooter? Where it happened, and how did you react to the situation?}
\item{Do you frequently encounter ``parked'' e-scooters while you walk?  Are there different experiences? }
\item{Do you feel unsafe having a parked e-scooter in the space you walk?}
\item{Any other experience with shared e-scooters you would like to explain?}
\item{Are you aware of the existing e-scooter regulations?}
        \end{enumerate}

        \subsubsection{Values}
        \begin{enumerate}
\item{What makes you choose walking over the other transport modes?}
\item{What do you expect as a good pedestrian experience? }
\item{Have you ever used a shared e-scooter? 
     If yes, what are the good things about e-scooters as you see? What are the bad things of e-scooter usage?
     If no, is there a reason for not using a shared e-scooter?}
\item{Do you think it would be helpful to have a system that recommends designated parking spots for e-scooters? As a pedestrian, what would you expect from such a system? }
\item{Do you think it would be helpful to have a system that recommends routes for e-scooters? As a pedestrian, what would you expect from such a system?}
        \end{enumerate}

 \subsection{Interview Questions for Cyclists}
        \subsubsection{Usage and frequency}
        \begin{enumerate}
\item{In a normal week, how often do you cycle in CBD?  
What areas/routes you cycle the most and in what days of the week and at what times of the day you used to cycle? }
        \end{enumerate}
        
        \subsubsection{Interactions}
        \begin{enumerate}
\item{Can you tell me about the last time you interacted with someone riding a shared e-scooter? Where did this happen, and what time of day was it? How did you react to the situation, and how did the e-scooter rider respond?}
\item{Do you often encounter shared e-scooter riders while you cycle in CBD? Is it the same experience always? }
\item{Can you tell me about the last time you interacted with a ``parked'' shared e-scooter? Where it happened, and how did you react to the situation?}
\item{Do you frequently encounter parked e-scooters while you cycle? Is it the same experience always?  }
\item{Any other experience with shared e-scooters you would like to explain? }
\item{Are you aware of the existing e-scooter regulations?}
        \end{enumerate}

        \subsubsection{Values}
        \begin{enumerate}
\item{What makes you choose cycling over the other transport modes?} 
\item{What do you expect as a good cycling experience? }
\item{Have you ever used a shared e-scooter? 
    If yes, what are the good things about e-scooters as you see? What are the bad things about e-scooter usage? 
    If no, is there a reason for not using a shared e-scooter? }
\item{Do you think it would be helpful to have a system that recommends designated parking spots for e-scooters? As a cyclist, what would you expect from such a system? }
\item{Do you think it would be helpful to have a system that recommends routes for e-scooters? As a cyclist, what would you expect from such a system?}
        \end{enumerate}
        
 \subsection{Interview Questions for Local Council Representatives}
        \begin{enumerate}
\item{What are the objectives of the shared e-scooter trial? }
\item{What were the considerations when planning the trial?}
\item{What are the benefits of shared e-scooters as you see? }
\item{What are the challenges/threats of shared e-scooters? }
\item{ Many e-scooter riders appreciate the convenience of being able to park e-scooters anywhere. However, this impacts on other road users, like pedestrians or cyclists? What are your thoughts on this?}
\item{What do you think about the infrastructure available for riding and parking?} 
\item{Do you think it would be helpful to have a system that recommends designated parking spots for e-scooters? As a local council representative, what would you expect from such a system? }
\item{Do you think it would be helpful to have a system that recommends routes for e-scooters? As a local council representative, what would you expect from such a system?}
        \end{enumerate}

 \subsection{Interview Questions for Service Provider Company Representatives}
       Thinking about the e-scooter trial in this city,
        \begin{enumerate}
\item{What are your objectives as a company? }
\item{What are the strategies you use to maximise the benefits? }
\item{What are the challenges you face during the e-scooter trial? }
\item{How do you overcome those challenges? }
\item{The rules for shared e-scooter riding vary from city to city. I'd like to hear your thoughts on two specific aspects in this city:
    What do you think about the current infrastructure allocated for e-scooter riding, particularly the rule that they shouldn't be ridden on footpaths? Do you feel this is effective, or could it be improved?
    What about the requirement for mandatory helmet usage? Do you think this rule is important? }
\item{Shared e-scooter parking rules also differ from city to city. What are your thoughts on allowing e-scooters to park anywhere, even not in designated no-parking zones?}
\item{What steps have you taken to minimise the negative impact of shared e-scooters on other road users?}
\item{Do you think it would be helpful to have a system that recommends designated parking spots for e-scooters? As a representative of e-scooter provider company, what would you expect from such a system? }
\item{Do you think it would be helpful to have a system that recommends routes for e-scooters? As a representative of e-scooter provider company, what would you expect from such a system?}
        \end{enumerate}

\end{document}